\documentclass[conference]{IEEEtran}
\usepackage{cite}
\usepackage{amsmath,amssymb,amsfonts}
\usepackage{algorithmic}
\usepackage{graphicx}
\usepackage{textcomp}
\usepackage{xcolor}
\usepackage{fancyhdr}

\usepackage[normalem]{ulem}
\usepackage{listings}
\usepackage{parcolumns}
\usepackage{tcolorbox}
\usepackage{diagbox}
\PassOptionsToPackage{hyphens}{url}
\usepackage[hyphens]{url}
\usepackage{hyperref}
\usepackage{multirow}
\usepackage{multicol}
\usepackage{microtype}
\usepackage{mathtools}
\usepackage{enumitem}
\usepackage[numbers,sort&compress]{natbib}
\usepackage{url}

\usepackage{balance}
\usepackage{siunitx}
\usepackage{optidef}
\usepackage{comment}

\usepackage[framemethod=tikz]{mdframed}
\usepackage{pifont}% http://ctan.org/pkg/pifont
\usepackage[noabbrev,capitalize]{cleveref}
\newcommand{\cmark}{\ding{51}}%
\newcommand{\xmark}{\ding{55}}%

\DeclareRobustCommand\encircle[1]{\tikz[baseline=(char.base)]{\node[shape=circle,fill,inner sep=0pt] (char) {\textcolor{white}{#1}}}}
\newcommand{\ignore}[1]{}
\usepackage{tikz}

\usepackage{xspace}
\let\svthefootnote\thefootnote

\makeatletter
\DeclareRobustCommand\onedot{\futurelet\@let@token\@onedot}
\def\@onedot{\ifx\@let@token.\else.\null\fi\xspace}

\newcommand{\red}[1]{\noindent{\color{red}{#1}}}
\newcommand{\blue}[1]{\noindent{\color{blue}{#1}}}

\definecolor{aliceblue}{rgb}{0.94, 0.97, 1.0}
\definecolor{darkgreen}{HTML}{006400}
\tcbset{width=1\columnwidth,boxrule=1pt,colback=aliceblue,arc=1pt,left=2pt,right=2pt,boxsep=0.5pt}
\newcommand\hl[1]{%
  \bgroup
  #1%
  \egroup
} 

\newcommand\shl[1]{%
  \bgroup
  #1%
  \egroup
} 

\definecolor{darkred}{rgb}{1.0, 0.1, 0.1}
\newcommand\mq[1]{%
  \bgroup
  #1%
  \egroup
}

\hypersetup{
    colorlinks=true,
    citecolor=violet,
    linkcolor=violet,
    filecolor=magenta,      
    urlcolor=violet
}

\def\BibTeX{{\rm B\kern-.05em{\sc i\kern-.025em b}\kern-.08em
    T\kern-.1667em\lower.7ex\hbox{E}\kern-.125emX}}

\title{Scalable and Secure Row-Swap: Efficient and Safe Row Hammer Mitigation in Memory Systems\vspace{-0.15in}}
\author{\IEEEauthorblockN{Jeonghyun Woo}
\IEEEauthorblockA{
University of British Columbia\\
jhwoo36@ece.ubc.ca}
\vspace{-0.4in}
\and
\IEEEauthorblockN{Gururaj Saileshwar$^{*}$}
\IEEEauthorblockA{
NVIDIA Research\\
gsaileshwar@nvidia.com}
\vspace{-0.4in}
\and
\IEEEauthorblockN{Prashant J. Nair}
\IEEEauthorblockA{
University of British Columbia\\
prashantnair@ece.ubc.ca}
\vspace{-0.4in}
}
\begin{document}
\maketitle
\thispagestyle{plain}
\pagestyle{plain}

\begin{abstract}
As Dynamic Random Access Memories (DRAM) scale, they are becoming increasingly susceptible to Row Hammer. By rapidly activating rows of DRAM cells (aggressor rows), attackers can exploit inter-cell interference through Row Hammer to flip bits in neighboring rows (victim rows). A recent work, called \emph{Randomized Row-Swap} (RRS), proposed proactively swapping aggressor rows with randomly selected rows before an aggressor row can cause Row Hammer. %RRS promises several years of protection against Row Hammer attacks.

Our paper observes that \emph{RRS} is neither secure nor scalable. We first propose the `Juggernaut attack pattern' that breaks \emph{RRS} in under \emph{1 day}. Juggernaut exploits the fact that the mitigative action of RRS, a \emph{swap} operation, can itself induce additional target row activations, defeating such a defense. Second, this paper proposes a new defense \emph{Secure Row-Swap} mechanism that avoids the additional activations from \emph{swap} (and \emph{unswap}) operations and protects against Juggernaut. Furthermore, this paper extends \emph{Secure Row-Swap} with attack detection to defend against even future attacks. While this provides better security, it also allows for securely reducing the frequency of swaps, thereby enabling \emph{Scalable and Secure Row-Swap}. The \emph{Scalable and Secure Row-Swap} mechanism provides years of Row Hammer protection with 3.3$\times$ lower storage overheads as compared to the RRS design. It incurs only a 0.7\% slowdown as compared to a not-secure baseline for a Row Hammer threshold of 1200.
\end{abstract}
 
%Make sure comparison against memory integrity in SGX etc. is in the intro.
 \ignore{
 -----------------------------------------------
 JH:
    Flow:
    1. Brief explanation of what RH is, and it's security issues (Why it is critical to solve).
        - Continuous DRAM manufacturing process scaling introduced a reliability issue such as RH.
        - Explain the security issue here (Cite a few RH-based attacks)
        - Mention that RH Threshold has reduced over the past years. This implies that RH will be more severe issue for future systems as DRAM scaling continues.
        - Thus, providing a secure and scalable RH protection is critical.
    
    2. Previous solutions and their limitations.
        - Many hardware-based solutions have been proposed, given that the seriousness of the problem.
        - Explain what victim-focused mitigation is & its limitations
            - Brief explanation of victim-focused mitigation
            - Limitations
                - Requires proprietary internal DRAM mapping information
                - Susceptible to newly introduced blast-radius related attacks such as Google's Half-double attack
                - Not scalable for future low RH threshold because
                    - cause severe performance overhead
                    - introduce significant HW overhead
        - Explain aggressor-focused mitigation
            - Recently, aggressor-focused mitigation (BlockHammer, RRS) has been proposed as a promising alternative to victim-focused mitigation, as it requires no proprietary information about DRAM and can protect the blast-radius related attacks. 
            - Unfortunately, both of previous aggressor-focused mechanisms are not scalable for future low RH threshold due to following limitations.
                - BlockHammer
                    - cause severe performance overhead
                    - introduce significant HW overhead
                - RRS
                    - Not secure anymore
                    - incurs significant performance overhead
                    - large HW overhead
    
    3. Our goal & our proposed idea (Brief explanation)
        - The goal of our paper is to enable a secure and scalable RH mitigation method even at future low RH Thresholds while working without proprietary internal DRAM mapping information. or The goal of our paper is to design a secure, low-cost, and high-performance RH mitigation method even at future low RH Thresholds while working without proprietary internal DRAM mapping information.
        - To this end, we propose 'Name", ~.
        - Brief high-level explanation of our idea.
    
    5~6/7: More detailed explanation of our proposed idea
        - TODO: Work on this just after having concrete design of our idea
    
    8. Evaluation
        - Highlight that our proposed idea has (1) low hardware (storage) overhead, (2) negligible performance overhead, (3) insignificant power overhead even in future low RH threshold (e.g., 1K) compared to other previous aggressor focused mitigation, BlockHammer & RRS.
    
    9. Overall contributions
 -----------------------------------------------
PN:
    Flow:
    1. Brief explanation of what RH is, and it's security issues (Why     it is critical to solve).
        - Continuous DRAM manufacturing process scaling introduced a reliability issue such as RH.
        - Explain the security issue here (Cite a few RH-based attacks)
        - Mention that RH Threshold has reduced over the past years. This implies that RH will be more severe issue for future systems as DRAM scaling continues.
        - Thus, providing a secure and scalable RH protection is critical.
        - While victim focused mitigation has been proposed, it has several pitfalls.
        - Thus, recent work -- RRS -- proposes aggressor-focused mitigation.
        - However, one can develop attacks to break RRS -- more so  as RH thresholds reduce.
        - We need scalable and secure aggressor-focused mitigation scheme.
    2.

 -----------------------------------------------

 }
\section{Introduction}\label{introduction}
Technology scaling has been a double-edged sword~\cite{archshield}. While it has enabled high-density Dynamic Random Access Memory (DRAM) chips, it has also uncovered security vulnerabilities. A key vulnerability called Row Hammer (RH)~\cite{kim2014architectural, kim2014flipping, moesi-prime, mutlu2022fundamentally} allows malicious processes to rapidly activate rows (aggressors) of DRAM cells and flip bits in their \emph{immediate} neighboring (victim) rows~\cite{aweke2016anvil, cojocar2019exploiting, frigo2020trrespass, gruss2018another, gruss2016rowhammer, kwong2020rambleed, seaborn2015exploiting}. 

There has been an arms race between RH attacks and defenses. To prevent RH, prior proposals tend to proactively refresh the contents of victim rows. This is called victim-focused mitigation (VFM)~\cite{frigo2020trrespass, kim2014architectural, kim2014flipping, lee2019twice, park_graphene:_2020}. 
However, new attack patterns, such as the \emph{half-double attack} from Google~\cite{half-double, half-double2}, have shown that they could trigger RH even in distance-of-2 (or more) rows away from the aggressor row by exploiting the mitigative action of VFM.
% However, new attack patterns, such as the \emph{half-double} attack from Google~\cite{half-double, half-double2}, discovered that the mitigative action of VFM, of refreshing neighbors, can be used to activate the neighbors and trigger RH in rows that are a distance-of-2 (or more) away from the aggressor row. 
%The half-double attack uses the victim-focused mitigating action (of refreshing rows that are a distance-of-1 away) to cause RH at rows that are at a distance-of-2 away. 
To overcome this, the state-of-the-art solution, \emph{Randomized Row-Swap} (RRS)~\cite{rrs}, uses an aggressor-focused mitigation mechanism. To this end, RRS swaps aggressor rows with random rows. Our paper finds that RRS is not secure. We show that, akin to the half-double attack, one can create a new access pattern by exploiting the mitigating action of RRS (the act of swapping rows) to break RRS. As a defense, our paper develops solutions that enables future-proof, secure, and scalable row swaps.
\let\thefootnote\relax\footnote{$^{*}$This work was partially performed when Gururaj Saileshwar was affiliated with Georgia Institute of Technology.}
\addtocounter{footnote}{-1}\let\thefootnote\svthefootnote

Malicious processes must activate their aggressor rows above a certain threshold, to trigger RH. This threshold is called the RH threshold (T$_{RH}$). The RH threshold must be crossed on a single row within an epoch of a refresh window (typically 64ms) to cause bit-flips within victim rows. To prevent this, RRS proactively swaps aggressor rows with randomly chosen rows before they reach T$_{RH}$. The number of activations at which a row is swapped is denoted by T$_{S}$, and the ratio of $T_{RH}$ to $T_{S}$ (\textit{i.e.}, $\frac{T_{RH}}{T_{S}}$) is called the `swap rate'. The choice of swap rate has security and performance implications. 

\begin{figure}[h!]
    \vspace{-0.1in}
    \centering
    \includegraphics[width=\columnwidth]{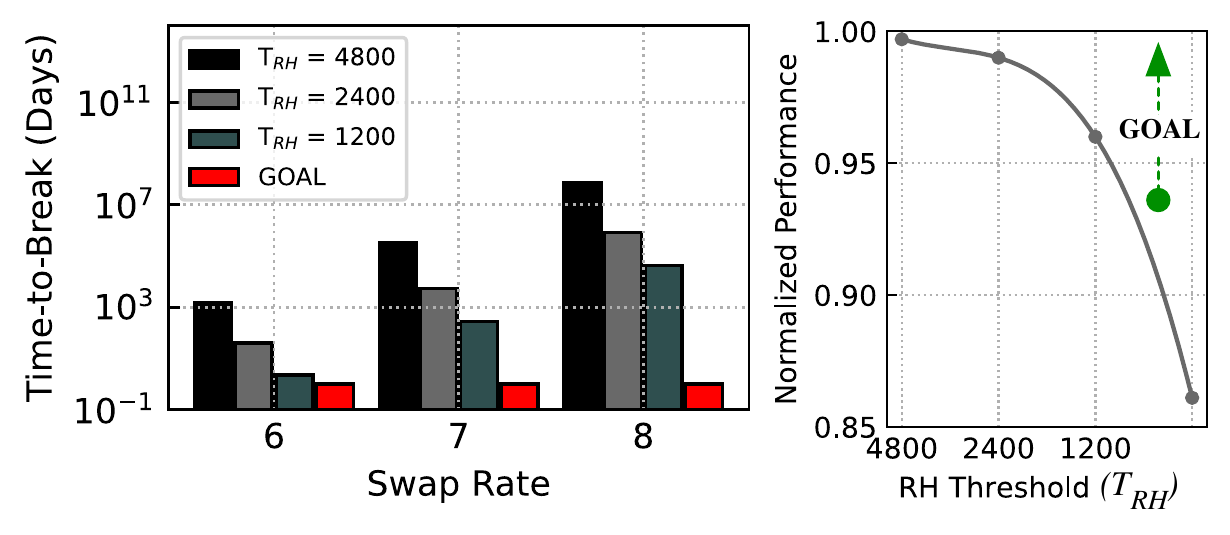}\vspace{-0.2in}
    \caption{(a) Time-to-break (in days) Randomized Row-Swap (RRS) with varying Swap Rate and Row Hammer Thresholds ($T_{RH}$). Our goal is to break RRS in under 1 day. (b) The normalized performance of RRS as values of $T_{RH}$ vary. Our goal is to minimize the performance overheads of RRS at lower values of $T_{RH}$ and enhance security; thereby making it scalable and secure.}
    \label{fig:goal}
\end{figure}

For security, the \emph{swap rate} is chosen such that no row in memory can reach the T$_{RH}$ number of activations within an epoch under years of attack. As shown in  Figure~\ref{fig:goal}, for a 32GB 16-bank DDR4-3200 system with a T$_{RH}$ of 4800 and a swap rate of 6 (default in RRS), it would take more than 10$^{3}$ days ($\sim$3 years) for an untargeted attack to succeed (as studied in RRS). A higher swap rate is even better for security, as it increases the attack time by increasing the adversarial effort of finding the attacked rows repeatedly.
So, our first goal is to investigate if a targeted attack pattern can break such defenses in under \emph{1 day}. Our second goal is to develop a secure defense against not just the Juggernaut attack pattern but even future \emph{unknown} attack patterns. %for solutions using randomized movements of rows as a mitigative action.

For performance, a lower \emph{swap rate} is better as this reduces the memory bandwidth and latency overheads. At a T$_{RH}$ of 4800 and a swap rate of 6, the system incurs an average slowdown of 0.3\% due to swaps. But as T$_{RH}$ drops in future generations (it has dropped 29$\times$ in 8 years~\cite{kim2014flipping,kim2020revisiting}), swaps will be needed after fewer activations, resulting in increased slowdowns and higher storage overheads to track more swapped rows.
So, our third goal is to enable a low-cost swap mechanism that securely tolerates lower swap rates to minimize performance and storage overheads.

%On the performance front, the swap rate needs to be minimized to reduce its memory bandwidth and latency overheads. As shown in Figure~\ref{fig:goal}, for a 32GB 16-bank DDR4-3200 system that has a T$_{RH}$ of 4800 and At a T$_{RH}$ of 4800 and swap rate of 6, the system incurs an average slowdown of \red{Y}\%. However, if we increase the \emph{swap rate} to 8, it increases the time-to-break for RRS to 10$^{13}$ days while also causing an average slowdown of Z\%. Our third goal is to develop a swap mechanism that minimizes performance overheads for swapping rows while also preventing the security pitfalls of RRS.

\ignore{
\begin{figure*}[t!]
    \centering
    \includegraphics[width=2\columnwidth]{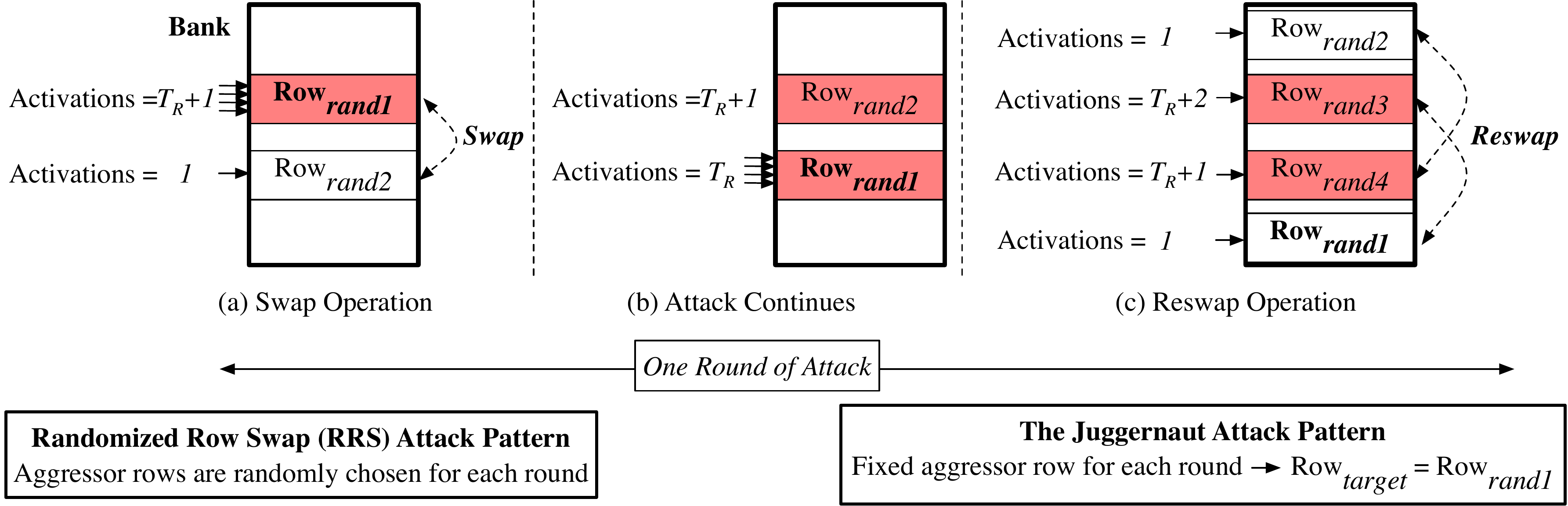}
    \caption{\emph{Swap} and \emph{Reswap} operations in Randomized Row-Swap (RRS). (a) If Row$_{rand1}$ is repeatedly activated, then after the first $T_{S}$ activations, it encounters a \emph{swap} operation. After additional $T_{S}$ activations, Row$_{rand1}$ will incur a \emph{reswap} operation. This increases the total activations at the original location of Row$_{rand1}$ to `$T_{S}$+2'. The attack pattern in RRS assumes that random rows are chosen in each round of attack. In contrast, the proposed Juggernaut attack pattern has a fixed aggressor row (Row$_{rand1}$ = Row$_{target}$) for each round of attack.}
    \label{fig:operation}
\end{figure*}
}

\noindent\textbf{Key Observation 1 -- Security}: The act of swapping rows, called the swap operation, itself incurs \emph{additional} row activations to read and write the original row. These additional activations can be used to bias any target row towards higher row activation counts. In the case of RRS, which picks random rows for swaps, let us assume that the internal chip address of the aggressor row is Row$_{aggr}$ and that of the randomly chosen row is Row$_{rand}$. A swap requires separately activating both Row$_{aggr}$ and Row$_{rand}$ and copying each row to the other's locations. Thus, if we repeatedly cause Row$_{aggr}$ to be swapped to new locations, then we can increase Row$_{aggr}$'s activation count each time due to the mitigating action (\textit{i.e.}, row swap)~\footnote{In spirit, this attack is inspired by the half-double attack against victim-focused mitigation. The half-double attack uses the mitigating act of refreshing neighboring rows to induce extra activations and trigger RH in farther rows.}.

After a swap, if we continue to activate Row$_{aggr}$ for other T$_{S}$ activations, the memory controller must first \emph{unswap} Row$_{aggr}$ before swapping it again with a newly chosen random location.
The \emph{unswap} operation itself also performs an additional row activation for Row$_{aggr}$.
Thus, we can develop a targeted attack that uses a combination of \emph{unswap-swap} operations on a single Row$_{aggr}$ to surpass the RH threshold (T$_{RH}$) within 64ms. For instance, even with 1 extra activation per \emph{unswap-swap}, up to 1700 activations are possible for a row within 64ms, purely due to \emph{unswap-swap} operations. This can significantly assist the demand activations, made to a row during an attack, to cross a T$_{RH}$ of 4800.
Such an attack, called the \emph{Juggernaut} attack, can break RRS in a significantly lower time ($<$1~day).

To defend against such attacks (and even future attacks), we propose \emph{Secure Row-Swap (SRS)}. SRS avoids \emph{unswap-swap} operations from biasing row activations and thereby protects against the Juggernaut attack. Moreover, we incorporate \emph{attack detection} in SRS to detect future attack patterns. As any successful attack requires swapping a single row multiple times, we deploy swap counters for mitigated rows in SRS to flag potential attacks. Thus, SRS enables attack-detection capability for protection against even future attacks.

\vspace{0.05in}
\noindent\textbf{Key Observation 2 -- Performance}: At lower RH thresholds (T$_{RH}$ $\leq$ 4800), even benign workloads tend to have frequently activated rows~\cite{moesi-prime}, requiring frequent swaps that can cause a slowdown. While reducing the swap rate (\textit{e.g.}, from 6 to 3) can reduce overheads and improve performance, this results in more frequent outlier rows that cross 3 swaps in an epoch (\textit{e.g.}, once every few hours), causing potential security breaches.
However, our swap-count-based attack detection mechanism can detect outlier rows, and then additional activations can be prevented by simply pinning these outliers in the Last Level Cache (LLC) for the rest of the refresh period (using $<$6\% of the LLC).
This enables extending SRS into a scalable design, called \emph{Scale-SRS}, that can employ a swap rate of 3 at lower values of T$_{RH}$, reducing the performance and storage costs.

\vspace{0.05in}
\noindent\textbf{Contributions}: This paper makes the following contributions. 
\begin{enumerate}[leftmargin=*]
\item We develop a new targeted attack pattern, \emph{Juggernaut}, that breaks RRS by exploiting the mitigative action of row swaps (and unswap operations) in under 1 day. 

\item We propose Secure Row-Swap (SRS), an RH mitigation that prevents \emph{unswap-swap} operations and defends against the Juggernaut attack pattern. Moreover, SRS also includes attack detection to detect future attack patterns against row-swap-based RH defenses.

\item We propose Scale-SRS, a scalable solution that can securely reduce the swap rate by combining outlier-based attack detection and LLC-pinning of outlier rows as mitigation. This improves the performance, storage costs, and scalability of row-swap-based RH defenses at lower T$_{RH}$ values.
\end{enumerate}

We show that Scale-SRS protects against the Juggernaut under reduced \emph{swap rates}. Compared to a baseline system that does not protect against RH, Scale-SRS incurs an average slowdown of only 0.7\%, even at the T$_{RH}$ of 1200. In a similar setup, we show that RRS can be broken in $<$1 day (regardless of the value of the swap rate), incurs a slowdown of $>$4\%, and has 3.3$\times$ higher on-chip storage overheads.
\ignore{JH
    -----------------------------Subsection Configurations--------------------------
    1. DRAM organization & Timing parameters (Can refer previous submissions)
    2. RowHammer & security implications (Can refer previous submissions): Include RH Threshold trend
    3. Threat Model (Can refer previous submissions)
    4. Victim-focused mitigation
        - Brief Explanations
        - Common limitations
            - Requires proprietary internal DRAM mapping
            - Vulnerable to newly introduced blast-radius related attacks such as Half-Double
        - CBT
            - Brief explanations
            - Limitations
                - High storage overhead: 304KB SRAM when Trh*=1000
                - Large performance overhead: causes a large # of unneccesary refreshes because refreshes are performed in groups of row (this cause high-latency also) that may include rows that are not aggressors. 
        - TWICE
            - Brief explanations
            - Limitations
                - Huge storage overhead: 2MB (0.9MB SRAM + 1.1MB CAM per rank) when Trh*=1000
                - Requires floating point operations (Incurs huge latency) when Trh* < 16K
        - Graphene
            - Brief explanations
            - Limitations
                - High storage overhead: 143KB CAM when Trh*=1000
    5. Aggressor-focused mitigation
        - BlockHammer
            - Brief Explanations
            - Limitations
                - Severe storage overhead: 229KB (184KB SRAM + 45KB CAM per rank) when Trh*=1000
                - High performance overhead
                    - Ex) When Trh*=1000, after 500 activations to the row, the row can be assessed every 128us
        - RRS
            - Brief Explanations
            - Limitations
                - Not secure
                    - Vulnerable to reswap-based attacks(TRH*=1000, swap threshold = 8)
                        - Swap operation itself generates additional 2 unavoidable activations. By repeatedly accessing the same row, attacker can cause the RowHammer to the row
                            - EX: 1 swap and 5232 reswap operations = 10589 ACTS > TRH*
                - Significant storage overhead (T*=1000, swap threshold = 8): 2.24MB SRAM
                - Severe perofrmance overhead (T*=1000, swap threshold = 8)
                    -Duty cycle becomes 0.48: Means available maximum # of activations in tFEW becomes half 
    6. Our goal
        - The goal of our paper is to enable a secure and scalable RH mitigation method even at future low RH Thresholds while working without proprietary internal DRAM mapping information. Or The goal of our paper is to design a secure, low-cost, and high-performance RH mitigation method even at future low RH Thresholds while working without proprietary internal DRAM mapping information.

}

\section{Background and Motivation}\label{background}
\subsection{Threat Model}\label{threat_model}
We assume a target system in which an Operating System (OS) provides process isolation using virtual memory and page tables. The memory system is composed of DRAM modules that are vulnerable to Row Hammer (RH). The attacker(s) run a malicious program in the \emph{user} privilege and activate DRAM rows rapidly. These rows, called the aggressor rows, can flip bits (by leaking charge) in their neighboring victim rows.

We assume that an attack succeeds if an aggressor row can trigger a bit-flip (\textit{i.e.}, when it incurs more activations than the RH Threshold (T$_{RH}$) within a refresh interval of 64ms). Similar to prior work, to showcase the effectiveness of our technique, we use a T$_{RH}$ value of 4800~\cite{rrs} (also lower T$_{RH}$ values to show scalability). It is the lowest demonstrated T$_{RH}$ value for any attack pattern, including Single-Sided~\cite{kim2014flipping}, Double-Sided~\cite{seaborn2015exploiting}, or Half-Double~\cite{half-double} attack patterns.

\subsection{Memory Organization and Timing Parameters}\label{dram_org}
A DRAM-based memory system consists of independent channels that are managed by individual memory controllers. Each channel consists of ranks which are composed of several banks that operate in parallel over a common memory bus. Each bank contains rows of DRAM cells that are accessed via a \emph{row-buffer}. The memory controller issues an activate (\emph{ACT}) command to bring data into the row-buffer. To access another row, the memory controller must replace the existing data in the row buffer by issuing the precharge (\emph{PRE}) command and subsequently issuing another ACT command. 

Each \emph{ACT} command leaks a small fraction of the charge within the DRAM cells of neighboring rows. DRAM cells also leak charge naturally and employ refresh operations (typically at 64ms intervals) to maintain data integrity. The time between consecutive ACT commands into the same bank is determined by the parameter  {\em $t_{RC}$} (Row Cycle Time). $t_{RC}$ is approximately 45ns for DDR4 systems. Thus, if we discount the time spent on refresh, a bank can experience up to 1.36 million activations ($ACT_{max}$) in the 64ms refresh window.

\subsection{Row Hammer (RH) Thresholds Over Time}\label{row_hammer}
The attacker(s) can use RH to flip bits in victim rows by activating an aggressor row above the RH Threshold (T$_{RH}$). To make matters worse, the value of T$_{RH}$ has reduced dramatically due to technology scaling. Table~\ref{table:RHT} shows the demonstrated values of T$_{RH}$ across different DRAM generations. The table uses {\em old} and {\em new} to distinguish different versions of the same standard that span multiple technology nodes. The value of T$_{RH}$ has reduced by nearly 29$\times$ in the last 8 years -- specifically from 139K~\cite{kim2014flipping} to 4.8K~\cite{kim2020revisiting}.

\begin{table}[!htb]
  \centering
  \begin{small}
  \caption{Row Hammer Threshold -- From 2014 to 2021}
  \label{table:RHT}
  \begin{tabular}{cc}
    \hline
    \textbf{DRAM Generation} & \textbf{RH-Threshold} \\ \hline %\hline
    DDR3 (old)     & 139K ~\cite{kim2014flipping} \\ %\hline
    DDR3 (new)     & 22.4K ~\cite{kim2020revisiting} \\ %\hline
    DDR4 (old)     & 17.5K ~\cite{kim2020revisiting} \\ %\hline    
    DDR4 (new)     & 10K ~\cite{kim2020revisiting} \\ %\hline \hline  
    LPDDR4 (old)   & 16.8K ~\cite{kim2020revisiting} \\ %\hline 
    LPDDR4 (new)   & 4.8K ~\cite{kim2020revisiting} - 9K~\cite{half-double} \\ \hline   
  \end{tabular}
  \end{small}
\end{table}

In practice, attackers have used RH to flip bits in the page table and cause privilege escalation~\cite{cojocar2019exploiting,frigo2020trrespass,gruss2018another, seaborn2015exploiting}. Attackers have also used RH to read confidential data~\cite{kwong2020rambleed}.

\subsection{Tracking Rows}~\label{trackers}
A key area of research has focused on developing efficient designs to track aggressor rows~\cite{park_graphene:_2020,lee2019twice,PROHIT}. Tracking aggressor rows helps issue timely mitigation. The row trackers could be placed within DRAM chips or memory controllers~\cite{frigo2020trrespass,mithril,hydra}. As the tracking mechanism is orthogonal to our mitigation mechanism, it is not our main focus. We evaluate our design with the state-of-the-art trackers, Hydra~\cite{hydra} and the Misra-Gries tracker (used in RRS~\cite{rrs} and Graphene~\cite{park_graphene:_2020}), although our mitigation is compatible with any aggressor tracker.

\subsection{Victim-Focused Mitigation}~\label{afm}
The victim-focused mitigation (VFM) refreshes the victim rows before the aggressor row receives more than T$_{RH}$ activations~\cite{park2020graphene,PROHIT,lee2019twice,mithril,kim2014flipping,kim2014architectural}. The number of victim rows near an aggressor row is determined by the {\em blast radius}~\cite{loughlin2021stop}. If the blast radius is n (where n $>$ 0), we would need to refresh \emph{n} rows on both sides of an aggressor row.

VFM tends to have two key concerns. 
First, VFM mechanisms implemented in the memory controller need to know the internal chip mappings of DRAM rows, specifically the set of neighboring rows for any row. Unfortunately, this proprietary internal row mapping information is not exposed to the memory controller~\cite{mithril}. Alternatively, VFM methods can be implemented inside DRAM chips, but this requires an additional interface to coordinate with the memory controller. Second, as shown by the recent half-double attack~\cite{half-double,half-double2}, refreshing \emph{n} victim rows can itself cause RH on the \emph{n+1}$^{th}$ victim row. To overcome this, recent proposals suggest using aggressor-focused mitigation. These proposals either blacklist the aggressor rows or break their spatial correlation with victim rows by displacing the aggressor rows~\cite{rrs,yauglikcci2021blockhammer,stefanrowhammer}. 
% First, it needs to know the internal chip mappings of DRAM rows, specifically the set of neighboring rows for any row.
% This is non-trivial as the internal chip addresses are scrambled.
% Therefore, VFM mechanisms are typically implemented inside DRAM chips and with the CPU-based memory controller only coordinating the mitigation. Ideally, it would be valuable for the memory controller to have \emph{full control} over the mitigation mechanism.

\thispagestyle{empty}

\subsection{Aggressor-Focused Mitigation: Randomized Row-Swap}~\label{rrs-mitigation} 
Randomized Row-Swap (RRS)~\cite{rrs} is the state-of-the-art aggressor-focused mitigation mechanism. RRS uses the memory controller to swap aggressor rows with randomly selected rows. The activation threshold for initiating a swap is typically much lower than T$_{RH}$ and is denoted by T$_{S}$. The fraction, $\frac{T_{RH}}{T_{S}}$, is called the \emph{swap rate}. Typically, the \emph{swap rate} is chosen such that RRS can tolerate several years of attacks. For instance, for a DRAM bank with 128K rows with a T$_{RH}$ of 4800, RRS, with a \emph{swap rate} of 6, can tolerate more than 3 years of attacks by an adversary continuously hammering randomly selected rows. This is because it is challenging for the attacker to guess the location of the aggressor rows as they are constantly shuffled. Since swaps impact both the security and performance of RRS, we dive deeper into its design.

%The swap rate determines the security and performance overheads of RRS -- higher the swap rate, better the security, but worse the performance.

\vspace{0.05in}
\textit{Swaps and Unswaps in RRS}:\label{swap-unswap}
RRS swaps a candidate row each time it crosses T$_{S}$ activations with a randomly chosen row (\emph{swap-partner}) within the bank. If the row needs to be swapped again in the same refresh window (due to T$_{S}$ more activations), the row and its current swap-partner need to be \emph{unswapped} before they may be swapped with new partners. 
%These operations have security and performance implications.

\begin{figure}[h!]
    \vspace{-0.1in}
    \centering
    \includegraphics[width=\columnwidth]{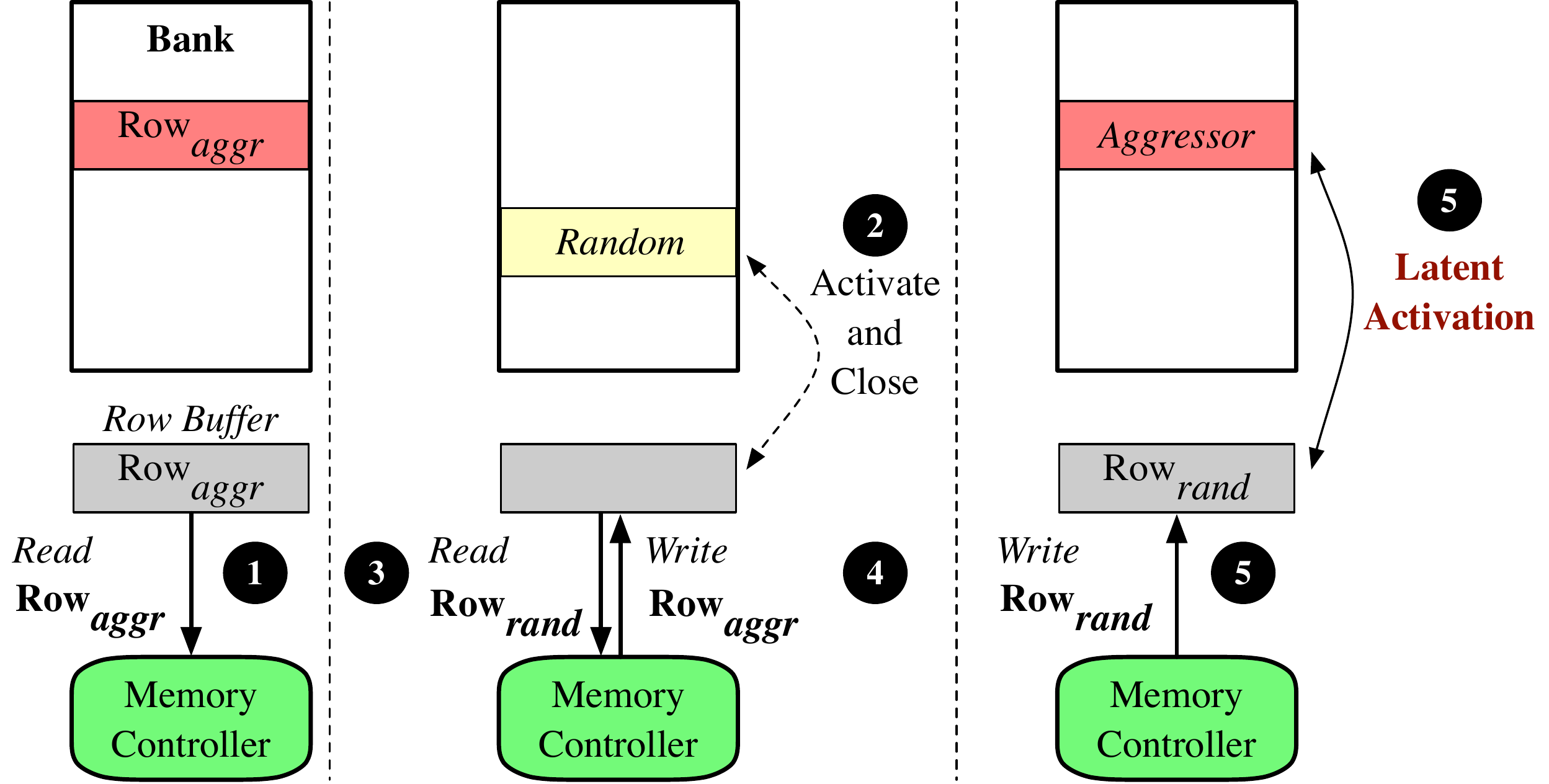}
    \caption{The latent activation on the aggressor row caused by a swap operation. This is primarily due to the fact that it takes five steps to activate two different rows (Row$_{aggr}$ and Row$_{rand}$) and thereby exchange their data contents.}
    \vspace{-0.1in}
    \label{fig:swap-latent}
\end{figure}

\vspace{0.05in}
\noindent\textbf{1. Security Implication}: Each mitigative action of \emph{swap} on an aggressor row itself causes one additional \emph{latent activation} at the original physical location of the aggressor row, which may be exploited by a new attack. To see why this occurs, consider the five steps in a swap operation, as shown in Figure~\ref{fig:swap-latent}:
\begin{enumerate}[leftmargin=*]
\item First, the aggressor row, Row$_{aggr}$ (activated by the attacker), is read out to the memory controller, as shown by \encircle{1}.
\item Then, the randomly chosen row, denoted as Row$_{rand}$, is activated, and this closes row Row$_{aggr}$, as shown by \encircle{2}.
\item The original data of Row$_{rand}$ is read out, as shown by \encircle{3}. 
\item The data of Row$_{aggr}$ is then written into the physical location of Row$_{rand}$, and that row is closed, as per \encircle{4}.
\item Finally, the original location of Row$_{aggr}$ is activated, and the data contents of Row$_{rand}$ are written into this location. This causes a latent activation, as shown by \encircle{5}.
\end{enumerate}

Thereafter, if any one of the pairs of swapped rows continues to receive T$_{S}$ activations, RRS would first unswap both these rows and then swap the aggressor row again to a new location.

\begin{figure}[h!]
    \centering
    \vspace{-0.1in}
    \includegraphics[width=\columnwidth]{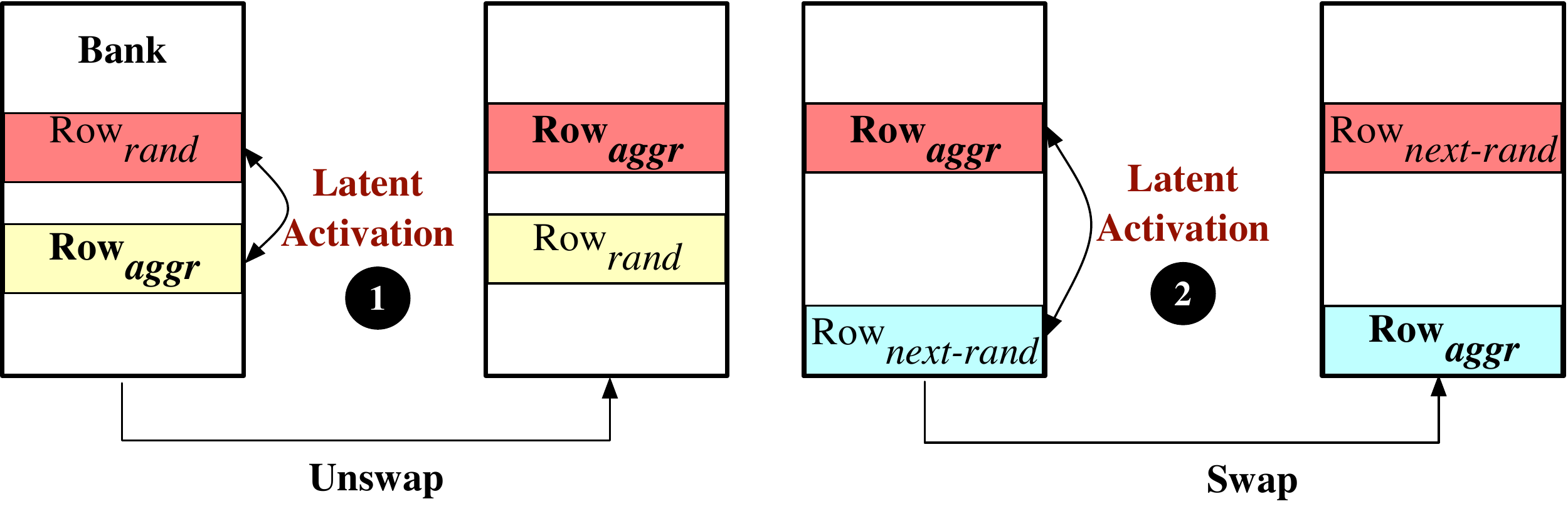}
    \vspace{-0.3in}
    \caption{Latent activations on the aggressor row caused by an unswap followed by a swap operation. These operations result in two additional activations.}
    \vspace{-0.1in}
    \label{fig:unswap-latent}
\end{figure}

All subsequent swaps for the aggressor row, within the refresh window, would be accompanied by an unswap, and together the unswap-swap operations cause up to \emph{two} latent activations at the aggressor's original physical row. 
As shown in Figure~\ref{fig:unswap-latent}, the first latent activation comes during the \emph{unswap}, which copies back the swapped aggressor row to its original location (as shown in \encircle{1}). Then the \emph{swap} of the aggressor row with the new location (Row$_{next\text{-}rand}$) also causes an additional activation to the aggressor's original location (as shown in \encircle{2}). Both steps incur extra activations because the row movements happen within the same bank, share a single row buffer, and require row-close and row-activate after each movement.

\ignore{
Subsequently, an unswap that accompanies the swap causes the aggressor row's location to incur up to \emph{two} latent activations. As shown in Figure~\ref{fig:unswap-latent}, this can be explained via two stages:
\begin{enumerate}[leftmargin=*]
\item During the unswap operation, the contents of the swapped rows are moved to their original locations. Similar to swap, as shown by \encircle{1} in Figure~\ref{fig:unswap-latent}, this would trigger \emph{one} latent activation in the original location of Row$_{aggr}$.
\item Thereafter, the contents of Row$_{aggr}$ swapped with another random row. As it is a swap operation, as shown by \encircle{2} in Figure~\ref{fig:unswap-latent}, Row$_{aggr}$ incurs \emph{one more} latent activation.
\end{enumerate}
}

Notably, if an attacker continuously activates the physical address of Row$_{aggr}$, its latent activations increases. In such a scenario, RRS issues mitigations that first cause one swap and then `$N$' \emph{unswap-swap} operations. Thus, the physical location originally storing Row$_{aggr}$ would have incurred up to $2N+1$ latent activations. This may be exploited by a new targeted attack to increase the activations for a location by exploiting latent activations from the mitigative operations.
%Namely, $2N$ latent activations from \emph{unswap-swap} operations and 1 latent activation from the original swap.

\begin{figure}[h!]
    % \vspace{-0.05in}
    \centering
    \vspace{-0.2in}
    \includegraphics[width=0.95\columnwidth]{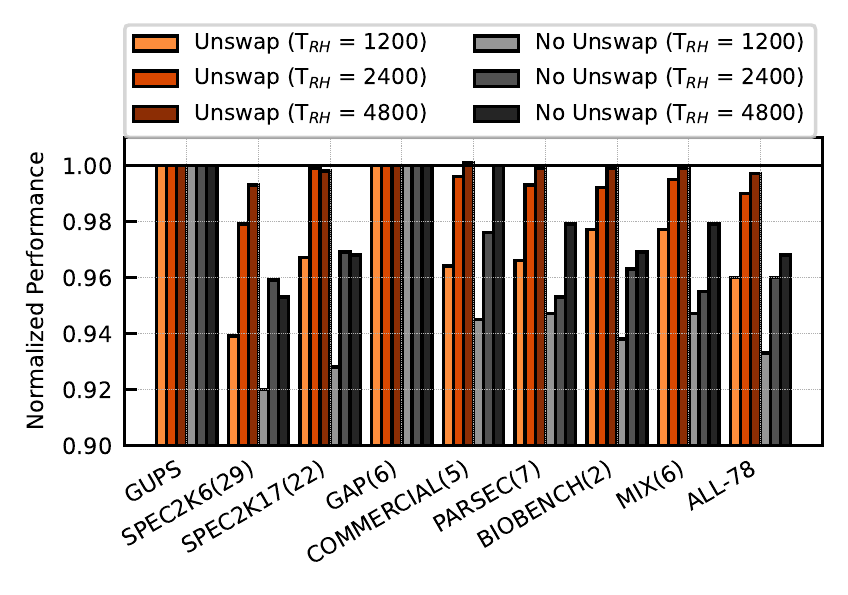}
    \vspace{-0.25in}
    \caption{The normalized performance of RRS, with and without immediate unswap operations, with respect to a baseline that does not mitigate against Row Hammer (RH). 
    %On average, not employing immediate unswap operations causes an average slowdown of up to 7\%.
    On average, not employing immediate unswap operations causes an additional slowdown of 3\% to 7\% at any given T$_{RH}$.}
    \label{fig:worstcase}
\end{figure}

\vspace{0.05in}
\noindent\textbf{2. Performance Implications from Swaps and Unswaps}: Unswap operations coupled with swaps are essential to ensure low-performance costs. This is because if an aggressor row is continuously swapped without first unswapping to its original location, it creates a chain of swapped rows that can introduce a large latency spike to unravel towards the end of a refresh interval. Figure~\ref{fig:worstcase} shows that if RRS does not employ immediate unswaps, it can cause an additional 3\% - 7\% slowdown on average compared to a design with immediate unswaps.

Consider a scenario in which Row$_{A}$ is swapped with Row$_{B}$. If Row$_{A}$ is continuously activated, it would need to be swapped again. Without the unswap, the new location containing Row$_{A}$ is now directly swapped with Row$_{C}$, and Row$_{A}$ is now in place of Row$_{C}$, while Row$_{C}$ is in place of Row$_{B}$, and so on. At the end of the refresh interval, all the swapped rows (Row$_{A}$, Row$_{B}$, Row$_{C}$, $\dots$) need to be placed back into their original locations. In practice, even one aggressor row can displace 1000s of random rows as it is swapped. Placing these random rows back together at the end of an epoch can cause a system to freeze up under hammering access patterns. 
Thus, designing a practical row swap mitigation without unswaps is non-trivial.

In the next section, we demonstrate how the latent activations of unswap-swaps can be exploited to break the defense and how a secure defense might be designed without unswap-swaps.
%To prevent this, the unswap that places the swapped row back to its original location is essential.

\thispagestyle{empty}

\ignore{

    <TODO>
    1. Analyze the attack scenario again when we assume that we could do swap with only one additional access to the row. --> Done
        - However, performing a swap with only one additional access won't be easy. This means that RRS has to be able to activate two rows in a bank concurrently, and it means we need to add more latches for DRAM subarrays similar to Subarray-level-parallelism (SALP). Also, since adding latch to each subarray is very expensive, so grouping a few subarrays and add latch for them is more feasible. However, in this case, the probability to choose the destination row that in the same subarray group will increase, which is the case that we have to perform 2 activations to the row.
            - Quantitative Analysis
                - Assume 512rows/subarray --> 256 subarrays per bank, 32subarrays/SA_Groups (8 SAG) --> 8 SAG
                --> The probability to choose the destination row in the same SAG: 12.5\%
        
    2. Think about a new scenario that can break the RRS that has an appropriate T to protect this issue. --> Done
    3. Figure out about the relationship between # of aggressor rows & breaking time

    Breaking Scenario: Just repeatedly access two rows (We need to have two rows to avoid rowbuffer hit).
        Concern: It seems like not a attack scenario, instead just a fundamental drawback of RRS for me.
        - Reason
            - Swap operation itself cause the additional 2 accesses for the selected rows (To Read the row, and to write the swapped row information there)
            - Thus, RRS will be broken deterministically when selected T cannot meet the equation below
                -> 
                1. Two activations for Swaps: TRH*/2 > T + 2*((A/2)/T)
                    - With TRH*=4.8K, RRS only can avoid this issue when 917<T<1483. However, if we choose T as 960, RRS will be broken only in 6.66 days.
                2. one activation for Swap: TRH*/2 > T + ((A/2)/T)
                    - With TRH*=4.8K --> 328 < T < 2072
                        -TODO: Think about how to break this now

        - Detailed steps
            1. Access a row pair (A,X) T times (Only explain about row X for the simplicity, Row A will go through the same process)
            2. RRS will swap the row X with the random row Y. 
                - It will incur additional 2 ACTs to X
            3. Access the row X T times again
            4. RRS will perform a reswap operation
                - Will swap the row X with the random row W
                    - Will cause additional 2 ACTs to X
                - Will swap the row Y with the random row Z
            5. Will go through step 3 & 4, repeatedly.
                - Results: Row X will received unwanted 2*((A/2)/T) accesses --> Total # of accesses to the row X = T (Expected # of accesses in RRs) + 2*((A/2)/T) (Unwanted Accesses). 
    Root cause of this issue?
        - RRS only allows 1:1 row mapping --> Reswap operations are necessary
            EX) Doesn't allow the mapping below. Only below case is allowed
                Src | Dest                       Src | Dest
                 X  |  Y                         X   | Y
                 Y  |  Z                         Y   | X
                 Z  |  X

    PN:
        B | C -- Phy B
        A | B -- Phy A
        C | B -- Phy C
        
}
\section{Juggernaut Attack Pattern} \label{sec::Juggernaut}
\subsection{Intuition and Overview}
The default attack studied in RRS employs a \emph{random-guess} strategy, where the attacker continuously picks random aggressor rows to activate and makes T$_{S}$ activations on it before it gets swapped. Eventually, the attacker hopes to repeatedly activate a single chip address by repeatedly guessing which row currently maps to it. For an RH threshold (T$_{RH}$) of 4800 and T$_{S}$ of 800, the attacker would need to correctly guess the mapping $\frac{4800}{800} = 6$ times -- essentially the \emph{swap rate}. This attack pattern exploits the \textit{birthday paradox} and takes years to break RRS. Rather than using \emph{only} the birthday paradox attack pattern, we develop a more effective attack pattern, \textit{Juggernaut}, that uses both latent activations and random guesses.

\begin{figure}[h!]
    \vspace{-0.05in}
    \centering
    \includegraphics[width=0.9\columnwidth]{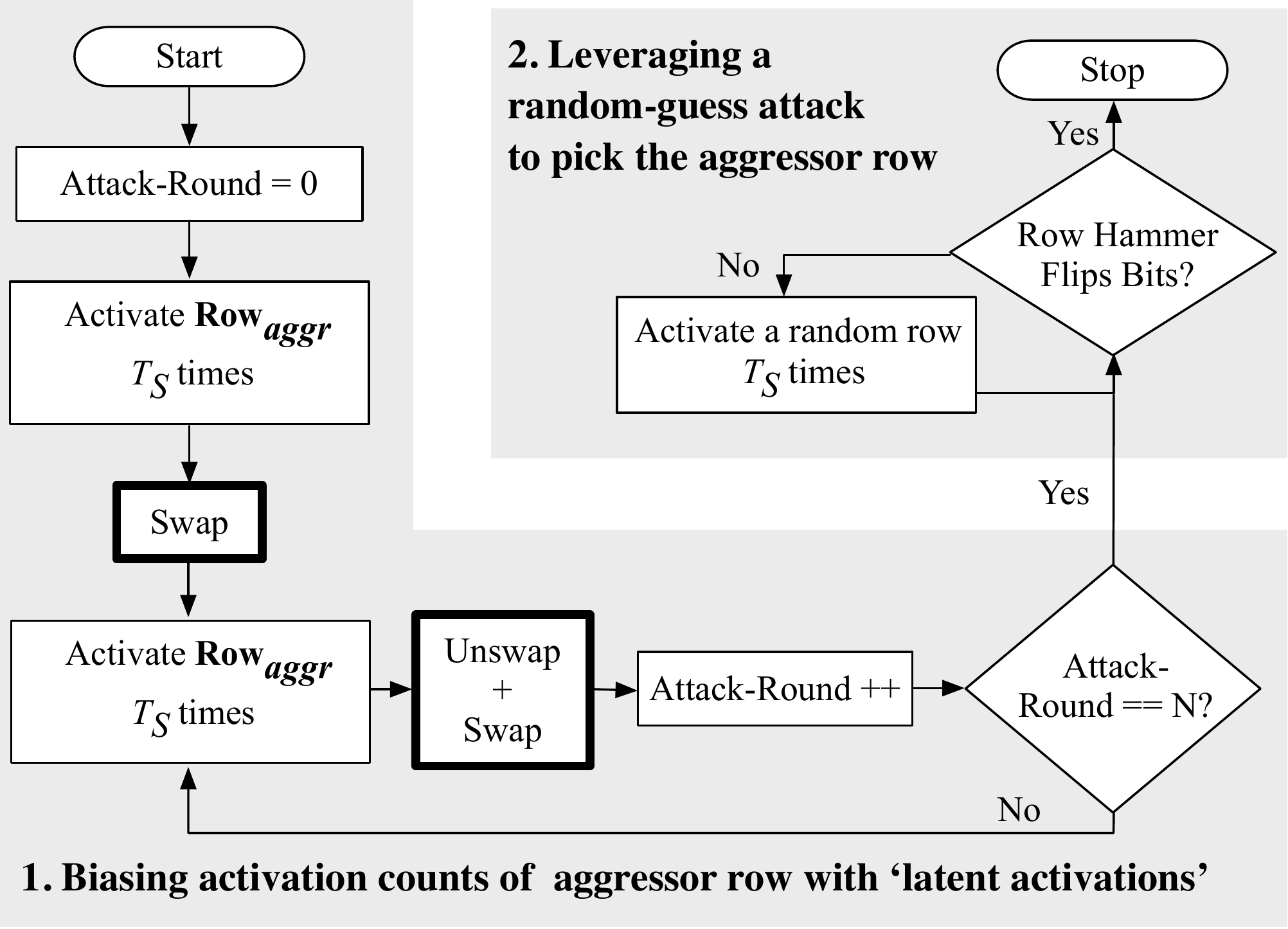}
    \caption{The high-level flow of the Juggernaut attack pattern. It consists of two parts. The first part biases an aggressor row with latent activations. The second part employs a random-guess attack.}
    \label{fig:overview}
    \vspace{-0.1in}
\end{figure} 

Figure~\ref{fig:overview} shows the high-level flow of Juggernaut (with latent activations and random guesses). Juggernaut uses latent activations to bias activations to a single chip address and thus reduces the adversarial effort for random guesses, as follows:

\begin{enumerate}[leftmargin=*]
\item First, we use latent activations to bias any one aggressor row towards a higher activation count. For instance, for a T$_{RH}$ of 4800, if the aggressor row incurs 800 unswap-swaps ($N$), then its original chip location would have incurred 1601 ($2N+1$) latent activations\footnote{Although a naive unswap-swap operation causes two latent activations, it is possible to optimize the unswap-swap using swap buffers in RRS (to be described in Section~\ref{sec:design}). In this case, depending on which row is selected first, a row gets either one or two additional latent activations. Thus, in this paper, we take an average of 1.5 latent activations per attack round.} (as described in Section~\ref{swap-unswap}). Additionally, it would have incurred T$_{S}$ (800) activations before its initial swap, and in total, 2401 activations. 

\item Subsequently, a \emph{random-guess} attack only needs to land T$_{S}$ (800) activations 3 times on the aggressor row for it to cross  T$_{RH}$ (4800) activations. As 3 is much lower than the \emph{swap rate} (6), it enables us to break RRS quickly.
\end{enumerate}

\ignore{
\begin{figure}[h!]
    \centering
    \includegraphics[width=\columnwidth]{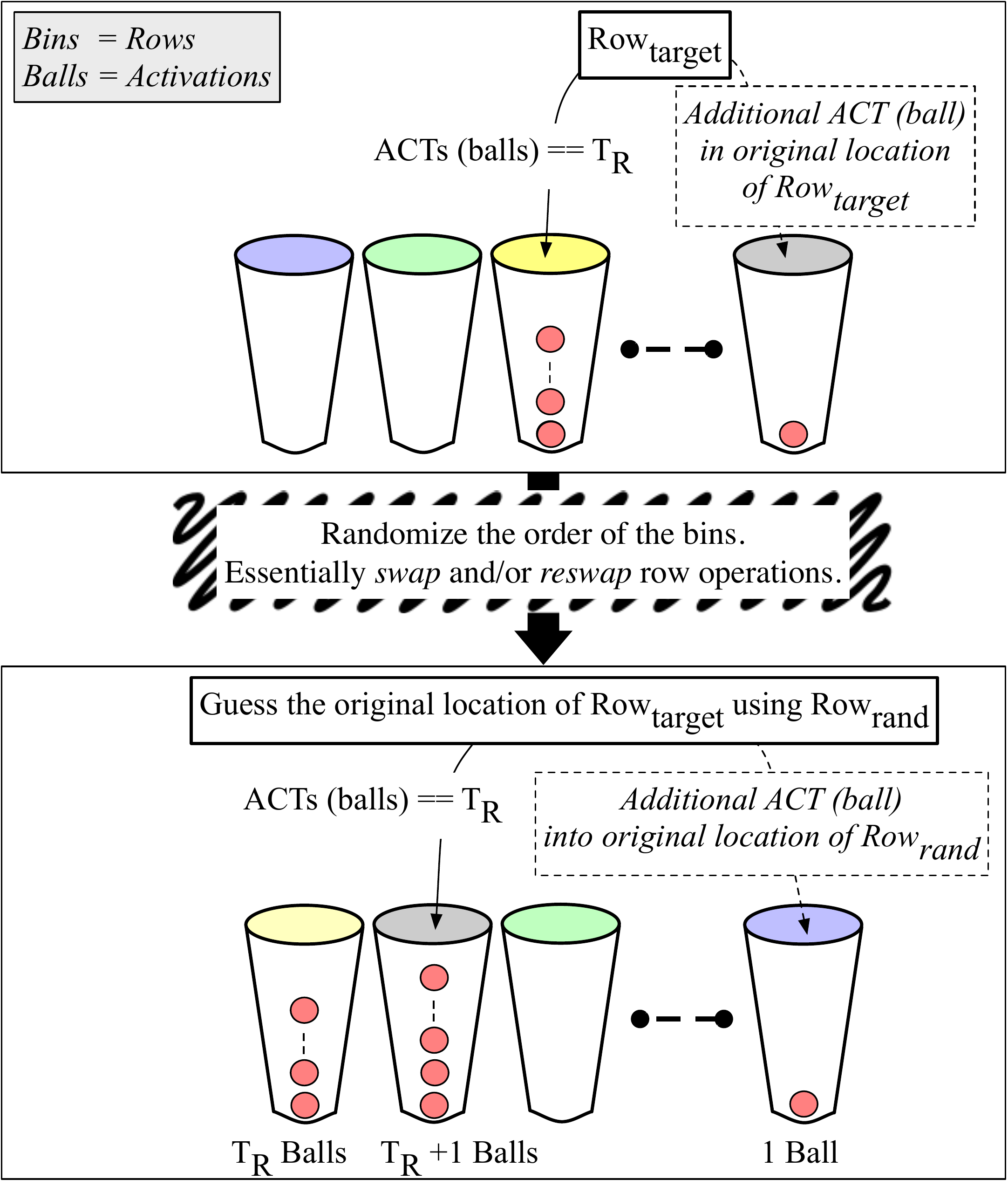}
    \caption{A bins and balls abstraction of the Juggernaut Attack Pattern. Balls represent the row activations and bin represents the rows. A \emph{swap} or a \emph{reswap} is denoted by changing the order of the bins.}
    \label{fig:bucketnballs}
\end{figure}
}
\ignore{
We explain the time-to-break RRS using such  as bins and balls analysis. Table~\ref{table:notations} shows the notations.

\begin{table}[h!]
\centering
\caption{List of Notations}
\resizebox{1\columnwidth}{!}{
\begin{tabular}{c|l}
 \hline
 \textbf{Notation} & \textbf{Description} \\
 \hline \hline
 \textbf{\emph{$T_{RH}$}} & Row Hammer (RH) threshold\\
 \textbf{\emph{R}} & Number of rows (buckets) in a bank\\
 %\hline
 \textbf{\emph{$T_{S}$}} & Activation threshold (number of balls) for remapping\\
 %\hline
 %
 \textbf{\emph{$N_{attack}$}} & Rounds of Attack to the Target Row ($Row_{target}$)\\
 \textbf{\emph{$N_{S}$}} & ACTs (ball) to $Row_{target}$ (target bucket) per \emph{reswap}\\
 %\hline
 \textbf{\emph{k}} & Swaps (bucket randomization) before RH is triggered\\
 \textbf{\emph{$t_{REFW}$}} & Refresh Time Window \\
 \textbf{\emph{$t_{RFC}$}}, \textbf{\emph{$t_{RC}$}} & Refresh Cycle, Row Cycle \\
 \textbf{\emph{$t_{reswap}$}} & Reswap Latency \\
 \textbf{\emph{$t_{swap}$}} & Swap Latency \\
 \textbf{\emph{$t_{ideal}$}} & Maximum time spent on ACTs within a refresh interval\\
 \textbf{\emph{$t_{rand}$}} & Time slots for randomly chosen rows attack-pattern \\
\hline
\end{tabular}
}
\label{table:notations}
\end{table}
}
\ignore{
JH: Explain this more in detail
}
\ignore{
As RRS uses a Misra-Gries tracker, the exact time of the refresh for specific rows cannot be known~\cite{park_graphene:_2020}. Thus, we can pick the original location of the target bin twice within each 64ms refresh interval. 
}

\subsection{Analytical Model of Juggernaut Attack Pattern}~\label{security}
We model our Juggernaut attack pattern statistically to better understand its impact. Table~\ref{tab:parameters_security} shows the parameters used in its analysis.
We also assume a memory controller with a closed-page policy similar to prior work~\cite{marazzi2022protrr}.
\begin{table}[h!]
    \vspace{-0.2in}
    \centering
    \caption{Key Parameters used in the Analytical Model} %{\color{blue}Will change numbers after confirming attack pattern, (Number) is currently used value}}
    \label{tab:parameters_security}
    \begin{tabular}{c|l}
    \hline
    \bf Parameter   &   \bf Definition  \\ \hline 
    $N$    &   Number of rounds of repeated unswap-swaps  \\
    $L$  &   Latent activations per round (up to 2)\\
    $G$   &   Number of Random Guess \\
    $R$   &   Number of Rows per Bank \\
    $t_{RC}$ &   Row Cycle Time \\
    $t_{reswap}$    &   Unswap-swap Latency = Reswap latency in RRS \\
    $t_{swap}$      &   Swap Latency \\ \hline
    \end{tabular}
    \vspace{-0.2in}
\end{table}
% \blue{  
% We showed that the naive implementation of unswap-swaps would make L as 2 in Section~\ref{swap-unswap}. However, it is possible to optimize the unswap-swap implementation using already implemented swap buffers in RRS to cause only one latent activation ($L$) for the target aggressor row (Row$_{aggr}$) sometimes.
% }% As an attacker can target a row across two refresh intervals, prior work have suggested using a swap-threshold that is 2$\times$ the designed threshold of T$_{S}$~\cite{park_graphene:_2020}.
\begin{tcolorbox}
\textbf{Goal:} For a successful RH attack, any aggressor row (Row$_{aggr}$) should incur $\geq$ T$_{RH}$ activations (ACTs).
\end{tcolorbox}

\noindent\textbf{1. Biasing an Aggressor Row with Latent Activations}\\ 
We consider $N$ attack rounds. Each round increases the latent activations of Row$_{aggr}$ by $L$ -- as shown in footnote 2, $L$ is 1.5. Furthermore, if the attack is timed precisely, an attacker can target a row $2 \times T_{S} - 1$ times before encountering an initial mitigative action (\textit{i.e.}, swap operation) that causes one latent activation. This exploits the fact that the refresh operations may not be synchronized with the reset of trackers~\cite{park2020graphene,hydra}. Equation~\ref{eqa:deterministic_accesses} shows the number of activations in the aggressor row (ACT$_{aggr}$) after $2 \times T_{S}$ initial activations, composed of $2 \times T_{S} -1$ direct activations and one latent activation, and $N$ rounds of latent activations ($L$).
\begin{equation}\label{eqa:deterministic_accesses}
   ACT_{aggr} =  2\times T_{S}+ (L \times N)
\end{equation}

\ignore{
\begin{table}[tb]
    \centering
    \caption{Parameters for Security Analysis} %{\color{blue}Will change numbers after confirming attack pattern, (Number) is currently used value}}
    \label{tab:parameters_security}
    \begin{tabular}{c|c|c}
    \hline
    \bf Parameter   &   \bf Definition                                  &  \bf Value  \\ \hline 
    $N_{attack}$    &   Number of repeated swap-unswaps             &  (1,067)  \\
    $N_S$           &   Number of latent activations per $N_attack$ &  (1.5)    \\
    B               &   Number of Random Guess                          & (437) \\
    N               &   Number of Rows per Bank                     &   131,072 \\
    $T_{RH}$        &   Row Hammer Threshold                        &   4.8K    \\
    $T_{S}$         &   Swap Threshold                              &   800     \\
    $t_{RC}$        &   Row Cycle Time                              &   45ns    \\
    $t_{reswap}$    &   Unswap-swap Latency (Reswap latency in RRS) &   (5.4us) \\
    $t_{swap}$      &   Swap Latency (Currently: 1 eviction + 1 swap) &   (2.7us)\\ \hline

    \end{tabular}
\end{table}
}

After $N$ rounds, the additional activations required for Row$_{aggr}$ to cause a bit flip are denoted as ACT$_{left}$ and can be represented using Equation~\ref{eqa:additional}.
\begin{equation}\label{eqa:additional}
    ACT_{left} = T_{RH} - ACT_{aggr}
\end{equation}

\noindent\textbf{2. Employing the Random-Guess Attack}\\
To further activate Row$_{aggr}$, as the attacker does not know its original location, they can repeatedly choose a random row (Row$_{rand}$) and activate it T$_{S}$ times. Some of these choices could land on the original location of Row$_{aggr}$. The number of \emph{swaps} ($k$) needed for this attack is denoted with Equation~\ref{eqa:k}.
\begin{equation}\label{eqa:k}
    k = \lceil \frac{ACT_{left}}{ T_S} \rceil
\end{equation}

$t_{RC}$ (45ns) is the minimum delay between activations. Let us assume a 64ms refresh interval (epoch). 
A DRAM bank performs 8192 refresh operations during an epoch, and each operation takes $t_{RFC}$ (350ns). Thus, only the remaining time the attacker can use ($t_{actual}$) is described by Equation~\ref{eqa:actual_attack time}.
\begin{equation}\label{eqa:actual_attack time}
    t_{actual} = 64ms - t_{RFC}\times8192
\end{equation}

In addition, the attacker has $N$ attack rounds ($t_{aggr}$) to bias the target aggressor row towards a higher activation count. As each attack round incurs T$_S$ activations to force an unswap-swap operation, with each unswap-swap operation incurring $t_{reswap}$ (5.4$\mu$s) latency\footnote{Note that, the Row Indirection Table (RIT) in RRS~\cite{rrs} evicts entries of the previous epoch \emph{before} the swap or unswap-swap operations (to be described in Section~\ref{sec:design}). To enable this, the attacker can fill RIT \emph{after} the first refresh interval (epoch).}, $t_{aggr}$ can be expressed by Equation~\ref{eqa:aggr_time}.
\begin{equation}\label{eqa:aggr_time}
    t_{aggr} = ((T_S-1)\times t_{RC} + t_{reswap})\times N
\end{equation}

The time the attacker spends to cause an initial swap should also be considered. As the attacker could generate $2\times T_S-1$ activations until to cause an initial swap with $t_{swap}$ (2.7$\mu$s) latency, the total time left ($t_{left}$) for employing the Random-Guess attack is denoted by Equation~\ref{eqa:attack_time_for_random_rows}.
\begin{equation}\label{eqa:attack_time_for_random_rows}
    t_{left} = t_{actual}-t_{aggr}-(t_{RC}\times (2\times T_S-1)+t_{swap})
\end{equation}

The total number of possible random guesses ($G$) within a refresh interval (epoch) is calculated using Equation~\ref{eqa:Balls}. Each randomly chosen row (Row$_{rand}$) is activated T$_S$ times. These rows only incur the initial swap ($t_{swap}$) latency. This is because most of these rows are picked only once.
% As these rows are seldom picked more than once, they mostly incur only the latency of the initial swap (t$_{swap}$).
\begin{equation}\label{eqa:Balls}
    G = \frac{t_{left}}{t_{RC} \times (T_{S}-1)+t_{swap}}
\end{equation}

Assuming a bank with $R$ (128K) rows, a row has a probability of $p=\frac{1}{R}$ of being selected. Thus, the probability ($p_{k,T_S}$) of a row having been selected $k$ times within $G$ random guesses is described by Equation~\ref{eqa:pkt}.
\begin{equation}\label{eqa:pkt}
    p_{k,T_S} = {}_G C_k \times p^{k} \times (1-p)^{(G - k)}
    % p_{k,T_S} = {}_G C_k \times (1/R)^{k} \times (1-1/R)^{(G - k)}
    % p_{k,T_S} = {}_G C_k \times (\red{\frac{1}{R}})^{k} \times (1-\red{\frac{1}{R}})^{(G - k)}
\end{equation}

% The expected number of rows ($R_k$) that have been selected $k$ times in a refresh interval is denoted by Equation~\ref{eqa:expected}.
% \begin{equation}\label{eqa:expected}
% R_k = R \times p_{k,T_S}
% \end{equation}

% The success probability of our attack pattern is the probability ($p_s$) of having picked the aggressor row when we choose $R_k$ rows. This is calculated using Equation~\ref{eqa:ps}.
% \begin{equation}\label{eqa:ps}
    % p_s = 1 - (1 - \frac{1}{R})^{R_k}
% \end{equation}

Since we only have a single target row, the expected number of iterations ($AT_{iter}$) and the time ($AT_{time}$) for a successful attack are represented by using Equation~\ref{Attackiteration} and Equation~\ref{Attacktime}.
\begin{align}
    AT_{iter} &= \frac{1}{p_{k,T_S}}\label{Attackiteration}\\
    AT_{time} &= 64ms \times AT_{iter}\label{Attacktime}
\end{align}
%The experimental results with 100,000 iterations of Monte-Carlo simulation closely match the mathematically derived values from Section~\ref{security}.
\thispagestyle{empty}
\subsection{Juggernaut: Determining the Attack Rounds}\label{sec:juggeranut-attack-round}
\begin{figure}[h!]
    \centering
    \includegraphics[width=0.95\columnwidth]{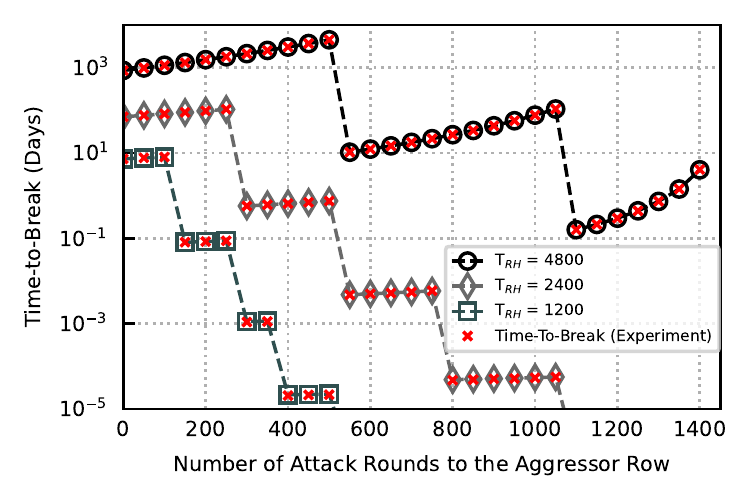}
    \vspace{-0.2in}
    \caption{Time-to-break RRS~\cite{rrs} with Juggernaut with varying attack rounds - both analytical and experimental results are shown. This analysis uses a swap rate of 6 for RRS. Juggernaut can break RRS in under 4 hours.}
    \label{fig:attacktime_vs_round}
    \vspace{-0.1in}
\end{figure}

Figure~\ref{fig:attacktime_vs_round} shows the time-to-break RRS with Juggernaut for different RH thresholds (T$_{RH}$) and varying rounds of attack. We also perform event-driven Monte Carlo simulations to validate our analytical model~\cite{faultsim1,faultsim2}. As shown in Figure~\ref{fig:attacktime_vs_round}, the results with 100,000 iterations of our Monte Carlo simulations closely match the values from our analytical model.

For a T$_{RH}$ of 4800, even after using a T$_{S}$ of 800 (swap rate of 6), Juggernaut takes only \emph{4 hours} to break RRS. 
% Furthermore, it can do this with $N = 1100$. 
In contrast, the naive attack pattern using only the birthday-paradox attack (used in RRS) takes $>$3 years to cause RH with T$_{S}$ of 800.

\begin{figure}[h!]
    \vspace{-0.1in}
    \centering
    \includegraphics[width=0.95\columnwidth]{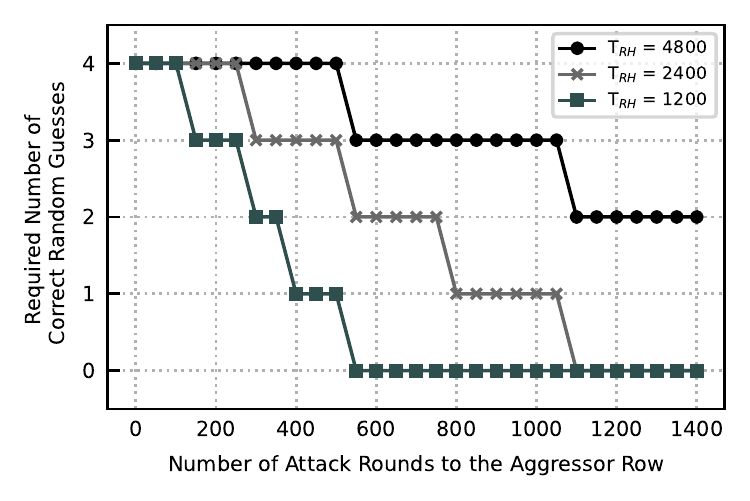}
    \vspace{-0.2in}
    \caption{The number of correct guesses required as the attack rounds vary. As the attack rounds increase, the attacker needs fewer guesses.}
    \label{fig:swaprate_vs_round}
\end{figure}

It is noteworthy to observe periodic `steep cliffs' in the time-to-break. This is because, as shown in Equation~\ref{eqa:k}, the value of $k$ (\emph{new swap rate}) is an integer. Thus, gradually varying the attack rounds can change the value of $k$ from one integer value to another -- which is manifested as a cliff in the time-to-break. 
Figure~\ref{fig:swaprate_vs_round} shows how the number of guesses required to break RRS ($k$) varies with attack rounds. As we increase the attack rounds, the attacker only needs fewer guesses. At a T$_{RH}$ of 4800, if the attacker uses $\leq500$ attack rounds, they would need to land at the original location of the aggressor row at least 4 times. In contrast, if we increase the attack rounds (say $\geq1100$), the attacker needs to guess the original location only twice.
Also, within the same required number of correct random guesses ($k$), we see that the time-to-break increases as the attack rounds increase, as shown in Figure~\ref{fig:attacktime_vs_round}. 
% However, within the same integer value of $k$, we see that the time-to-break increases as the attack rounds increase. 
This is because a larger number of attack rounds decreases the number of guesses ($G$) in Equation~\ref{eqa:Balls}.

Hence, we pick the number of attack rounds ($N$) such that it minimizes the value of $k$, while also maximizing the number of guesses ($G$). For instance, at a T$_{RH}$ of 4800, selecting $N$ as 1100 shows the best attack performance -- breaking RRS in under 4 hours. It is noteworthy to mention that, as shown in Figure~\ref{fig:swaprate_vs_round}, Juggernaut can break RRS in just 1 refresh period (64ms) using only the latent activations (unswap-swaps) at lower T$_{RH}$ values (\textit{e.g.}, 2400 and 1200). To make matters worse, the T$_{RH}$ value is highly likely to drop further due to the DRAM technology scaling -- T$_{RH}$ has already dropped by 29$\times$ from 2014 to 2022. Thus, it is vital to develop a low-cost protection technique not only against the Juggernaut attack but also other unknown attack patterns.

We also analyze a multiple-bank attack, where the attacker targets multiple banks instead of a single bank. However, such an approach considerably reduces the attack effectiveness. This is because it significantly decreases the number of possible activations in one refresh interval due to bank-to-bank activation delays and row migration latencies~\cite{rrs}. For instance, at a T$_{RH}$ of 4800 with a swap rate of 6, targeting all (16) banks in a channel increases the attack time from 4 hours to 9.9 years. Thus, we only focus on a single bank attack.

% \subsection{Juggernaut: Calibrating the Number of Guesses}
% Figure~\ref{fig:swaprate_vs_round} shows the required number of guesses to break RRS. As we increase the attack rounds, the attacker only needs fewer guesses. At the $T_{RH}$ of 4800, if the attacker use $\leq500$ attack rounds, they would need to land at the original location of the aggressor row at least 4 times. On the other hand, if we increase the attack rounds (say $\geq1100$), the attacker needs to guess the original location only twice.

\ignore{
\subsection{Changing the Swap (Remapping) Rate}
One can increase the swap rate to increase the security of RRS for a random attack-pattern. 
Figure~\ref{fig:attacktime_swaprate} shows the time-to-break for different values of $T_{RH}$ at varying swap rates.

For a random attack-pattern, as one increases the swap rate, the number of guesses required by the attacker increases, and this increases the time-to-break. On the contrary, the Juggernaut attack-pattern is resistant to higher swap rates as each swap increases the additional activations at the original location of the target row.
\begin{figure}[h!]
    \centering
    \includegraphics[width=1\columnwidth]{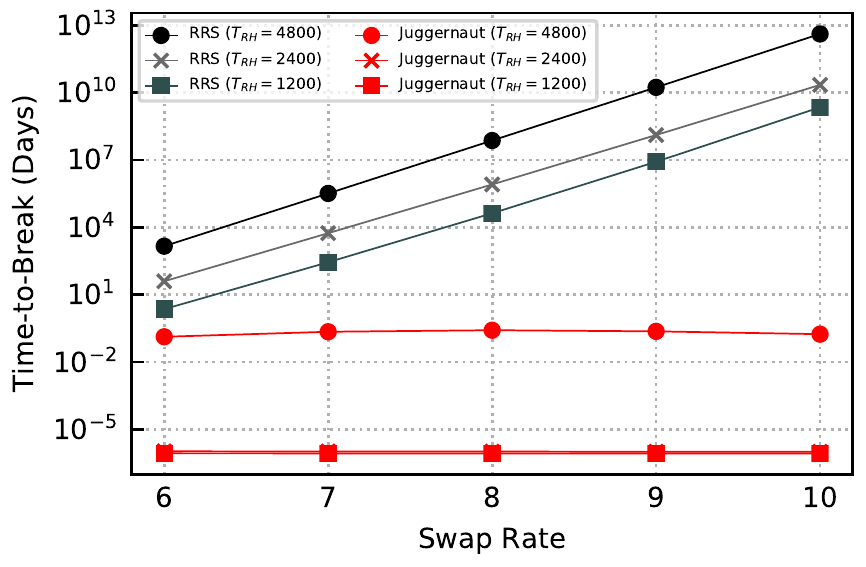}
    \caption{Time-to-break comparison between the random attack-pattern in RRS and the Juggernaut attack-pattern. The Juggernaut attack-pattern is resistant to increased swap rates and can break RRS in \emph{under 1 day}.}
    \label{fig:attacktime_swaprate}
\end{figure}

For $T_{RH}$ of 4.8K and a swap rate of 10, the random attack-pattern requires 10$^{13}$ days to break RRS -- greater than the age of the universe. On the other hand, Juggernaut can break this system in \emph{under 1 day}. This showcases the potency of Juggernaut. 

Generally, increasing the swap rate is not performance efficient. At higher swap rates, the memory controller will move rows even more frequently -- increasing the bandwidth and latency overheads of the memory system.
}
\ignore{
\subsection{Crafting Juggernaut}

{\color{blue}
Breif explnation by JH
\begin{itemize}
    \item Required time to break decreases abruptly at some points, then increases. This trend can be explained with Figure~\ref{fig:swaprate_vs_round}. Figure~\ref{fig:swaprate_vs_round} exhibits the changes in the required number of swaps to cause RowHammer (k) depending on the number of rounds-of-attack to the target row. Whenever k decreases, required time to break declines significantly, but for the same k increasing $N_{attack}$ diminishes the attack efficiency. This is because the attacker will have less opportunities to guess a swapped target row location (less balls to throw). Thus, the best attack pattern will be accessing the target row only to make k as lowest possible number, then try random guess for other available activations.
    
    \item We also analyze the attack-pattern using multiple target rows. Albeit this would allow the adversary great chances to guess target rows randomly, attacking multiple target rows reduce the maximum number of available of accesses to each target row. As a result, best reachable minimum k' becomes higher than single target row attack, which means the it will take much longer time to break RRS. Thus, we focus on the single target row attack.
    
    \item We may want to mention the multi-bank attack pattern also. The reason why multi-bank attack is worse than single-bank attack is because of reduced duty cycle. Specifically, since we share the swap buffer across the rank we must block the memory accesses to DRAM during the time, this will hurt the duty cycle hugely, and it will reduce the maximum number of available accesses to the target row also. Consequently, our best reachable minimum k becomes higher than single bank attack. 
\end{itemize}
}

\subsection{Results}
{\color{blue}
Will show security analysis results here.
}

\subsubsection{RRS with a Higher Swap Rate}
{\color{blue}
Will show that using higher k to provide a higher level of security still makes RRS much vulnerable to our Juggernaut attack.
Show security analysis results with Graph here.
Explain the reason intuitively: Smaller T enables Juggernaut to perform much more number of deterministic accesses.
}
}

\ignore{
<TODO>
    Think about better method to solve the security issue.

    Information for Secure and Scalable Row-Swap
    1. Overall Operations
        - High-level idea: Perform swap operations until the ACT_CNT < (TRH*/2)-1, 
                            and blacklist the row until the end of tREFW since ACT_CNT == (TRH*/2) - 1
            - Things to consider
                - Need to increase the access counter by 2 for each swap operation to avoid potential RH errors
        - Questions that need to answer for the operations
            1. How to track the actual # of accesses to the physical row
                - If we only add to the row into the priority queue when ACT_CNT > T, then how can we track T for each accessed row? Do we need to have another structure for this?
            2. How can we select the row to swap randomly?
                - If we swap the row with the row below, does that mean swaps not happen randomly?
                    ex) Access 0x0010 T Times --> Swap 0x0010 w/ 0x 0011?
                        0x0010 and 0x0011 will not be adjacent in DRAM, but anyway we can access 0x0010 by accessing 0x0011.
    2. Area (Storage) Overhead
        - # of entries for Priority Queue: A (MAX # of ACTs) / T (SCRS Threshold) 
            - Ex) TRH*=1000, k=4 --> 1360K/250 = 5440 entries
        - Entry Configuration: [Address, Swap_Count (k)]
            - Ex) [17-bits, 2-bits] = 19-bits
        - Estimated Overhead per bank (KB) = 12.62KB (Bigger than Graphene (8.97KB))
    3. Performance Overhead
        - 
    
}

\section{Mitigating Juggernaut with Secure Row-Swap}\label{sec:design}
\subsection{Overview and Intuition}
Secure Row-Swap (SRS) leverages the observation that latent row activations are due to the subsequent unswap and swap (\emph{unswap-swap}) operations. As latent activations are key to the success of Juggernaut, SRS prevents latent activations by avoiding \emph{unswap-swap} operations.

SRS observes that \emph{unswap-swap} operations create pairs of tuples of row mappings. This implies that if Row$_{A}$ maps to Row$_{X}$, then Row$_{X}$ also maps to Row$_{A}$. The pairs of tuples of mappings enable RRS to immediately unswap these rows.

Unlike RRS, SRS manages row mappings such that it can only employ the swap operation. For instance, in SRS, if Row$_{A}$ is repeatedly activated T$_{S}$ times, it will perform a swap operation by choosing a random row (say Row$_{Z}$), thereby destroying the original tuple pair. SRS is designed to lazily unswap rows (across epochs) into their original locations by using a small per-bank place-back buffer. The lazy unswap operations help mitigate performance overheads.

\subsection{Row Indirection Table}
The Row Indirection Table (RIT) tracks row remappings in RRS. SRS also uses a modified RIT. RIT is constructed as a Collision Avoidance Table (CAT)~\cite{MIRAGE}. The total number of entries in RIT (RIT$_{entries}$) depends on T$_{S}$ and the maximum number of activations (ACT$_{max}$) in a refresh interval (epoch). Additionally, the CAT structure is over-provisioned to prevent collision-based attacks~\cite{MIRAGE,rrs}. Furthermore, RRS stores RIT entries as tuples to enable efficient \emph{unswap-swap} operations.

For instance, if Row$_{A}$ and Row$_{B}$ are swapped, the RIT will have the tuples $<A,B>$ and $<B,A>$. If either Row$_{A}$ or Row$_{B}$ gets additional $T_S$ activations, both rows are unswapped and swapped. Assuming Row$_{A}$ is swapped with Row$_{C}$ and Row$_{B}$ is swapped with Row$_{D}$, then the RIT will now have the tuples $<A,C>$, $<C,A>$, $<B,D>$, and $<D,B>$. %The number of entries in RIT (as derived from RRS) is given by Equation~\ref{eqa:rit}.
%\begin{equation}\label{eqa:rit}
%   R_{entries} =  4\times\frac{ACT_{max}}{T_{S}}
%\end{equation}

% The maximum number of valid entries within the RIT for each epoch, called R$_{epoch}$, is calculated as Equation~\ref{eqa:ritmax}.
% \begin{equation}\label{eqa:ritmax}
%   R_{epoch} =  2\times\frac{ACT_{max}}{T_{S}}
% \end{equation}
% and \emph{unswap-swap} operations create two tuples.
% Furthermore, RIT stores entries as pairs -- thereby requiring an additional 2$\times$ more entries as compared to the original CAT design.
\ignore{
\begin{figure*}[t!]
    \centering
    \includegraphics[width=2\columnwidth]{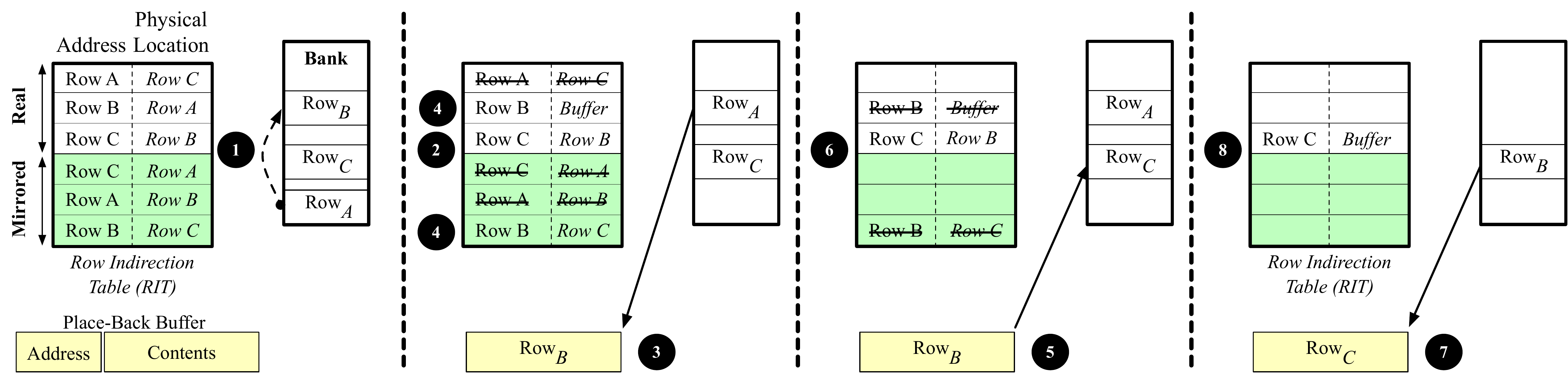}
    \caption{An overview of the \emph{place-back} operation and \emph{place-back} buffer for enabling Secure Row-Swap (SRS). SRS does not require tuples of row addresses in RIT. The \emph{place-back} buffer helps lazily store the rows that are displaced from the original location.}
    \label{fig:srs-swapback}
    \vspace{-0.2in}
\end{figure*}
}
\begin{figure*}[h!]
    \centering
    \includegraphics[width=2\columnwidth]{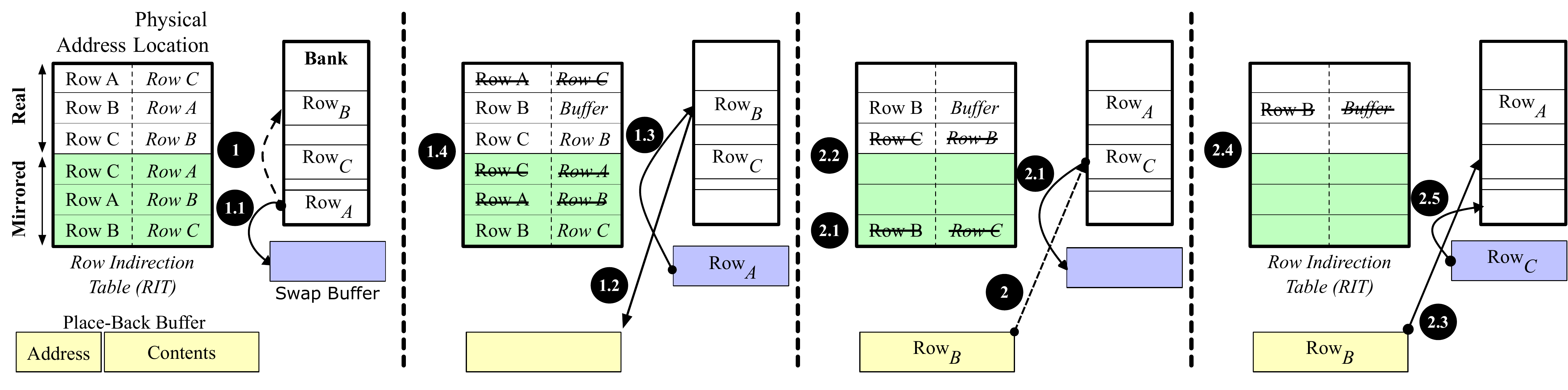}
    \caption{An overview of the \emph{place-back} operation and \emph{place-back} buffer for enabling Secure Row-Swap (SRS). SRS does not require tuples of row addresses in RIT. The \emph{place-back} buffer helps lazily store the rows that are displaced from the original location.}
    \label{fig:srs-swapback-v2}
    \vspace{-0.2in}
\end{figure*}
%We explain the RIT design for $T_{RH}$ of 4.8K, $T_{S}$ of 800, and ACT$_{max}$ of 1.36 million. RIT has 3400 tuples or 6800 entries (from Equation~\ref{eqa:rit}).
A lock bit is set for both tuple entries when they are brought into RIT. The lock bits are reset at the end of the epoch. RIT randomly evicts tuples from the previous epoch to insert new tuples. RIT uses lock bits to identify if the tuples are indeed from the previous epoch. 

% The tuples with their lock bits reset are lazily evicted out of the RIT. The SRS mechanism piggy-backs on the RIT design.

\subsection{SRS: Swap-Only Row Indirection}
SRS splits the RIT into two equal parts, namely, the real part and the mirrored part. Cumulatively, they have the same size as the RIT from RRS and retain the properties of CAT. The original mappings are stored in the real part, and the reverse mappings are stored in the mirrored part of the RIT.

\vspace{0.05in}
\noindent\textbf{1. Initial Swap}: Let us assume that Row$_{A}$ swaps with Row$_{B}$. The original RIT now contains the tuples $<A,B>$ and $<B,A>$. The mirrored RIT contains the tuples $<B,A>$ (for $<A,B>$) and $<A,B>$ (for $<B,A>$).

\vspace{0.05in}
\noindent\textbf{2. Subsequent Swaps}: Thereafter, if Row$_{A}$ receives T$_{S}$ activations again, then Row$_{A}$ is simply \emph{swapped} again -- \emph{without unswapping}. Let us assume Row$_{A}$ now swaps with Row$_{C}$. The $<A,B>$ entry in the original RIT is now updated to $<A,C>$. Additionally, as Row$_{C}$ is now placed in the original location of Row$_{B}$, a new $<C,B>$ is also added. However, the original RIT still maintains the valid entry $<B,A>$.

\begin{figure}[h!]
    \vspace{-0.1in}
    \centering
    \includegraphics[width=1\columnwidth]{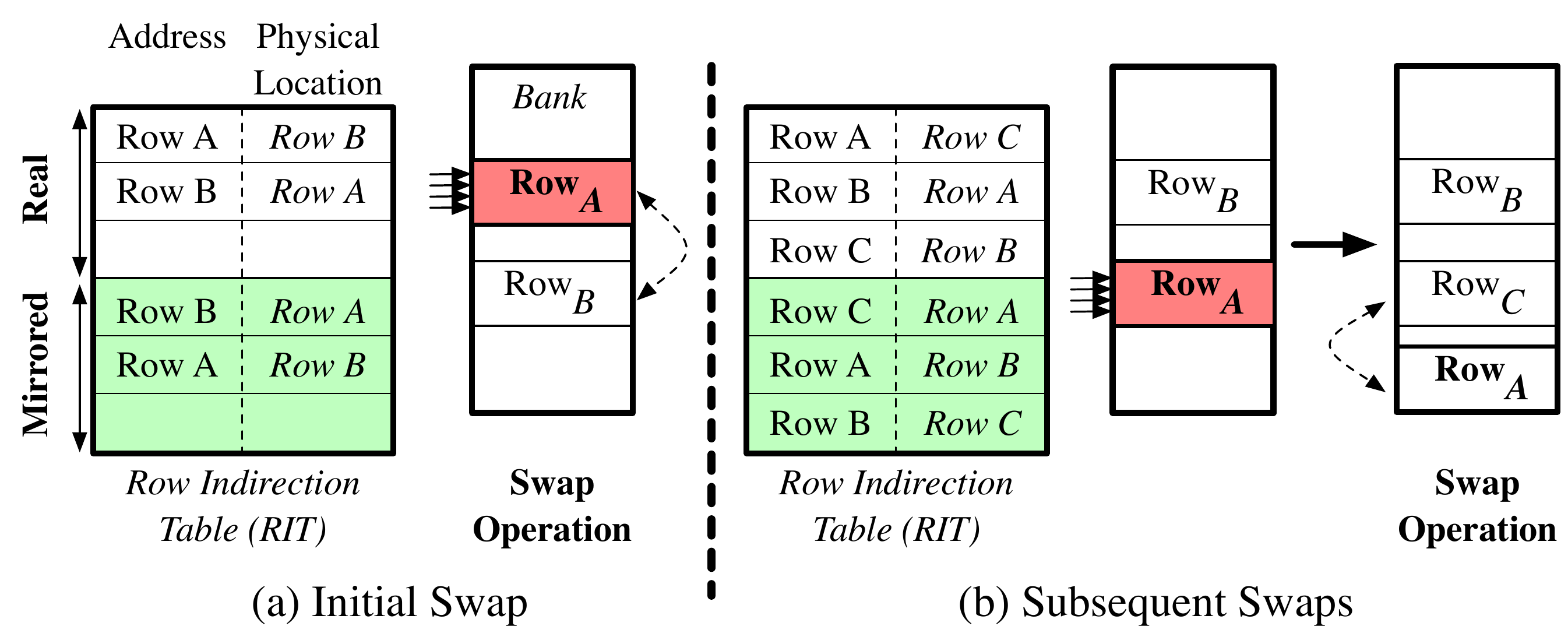}
    \caption{The RIT (real and mirrored) provides indirections to the rows involved in the \emph{swap} operations in SRS. The tuples in SRS do not have fixed pairs.}
    \label{fig:srs-rit}
    \vspace{-0.1in}
\end{figure}

The mirrored RIT is also updated with the reverse mappings of the entries in real RIT. Therefore, the mirrored RIT now contains $<C,A>$, $<A,B>$, and $<B,C>$. Figure~\ref{fig:srs-rit} shows these row mappings. A key difference between SRS and RRS is that the RIT tuples in SRS do not have fixed pairs. \emph{As there are no unswap operations, there is no latent row activation on the original location of the swapped rows}.

\subsection{SRS: Lazy Evictions and the Place-Back Buffer}
SRS employs lazy evictions of RIT entries from the previous epoch. These lazy evictions occur periodically in the current epoch. This design serves two purposes. First, the lazy evictions create space in the RIT for new entries for the next epoch. Second, due to their lazy nature, these evictions mitigate latency spikes as they are spread across the entire epoch.

SRS uses a per-bank `place-back' buffer that holds the contents of the rows that are being evicted. Consider a scenario where RIT is performing lazy evictions for the entries of the previous epoch. If the RIT has 1700 valid entries from the previous epoch, each valid entry will be lazily evicted periodically at the rate of $\frac{Epoch_{Time}}{1700}$ (\textit{i.e.}, $\frac{64ms}{1700}$). Note that, similar to RRS, the RIT is designed as a CAT. Thus, it can never be fully occupied and is resilient to conflict-based attacks.
\ignore{
As shown in Figure~\ref{fig:srs-swapback}, let us assume that the RIT contains mappings for Row$_{A}$, Row$_{B}$, and Row$_{C}$. If Row$_{A}$ is lazily evicted from the RIT, as shown by \encircle{1}, it will be moved into its original location by using the swap-buffers (already present in RRS). As shown by \encircle{2}, the RIT also invalidates the entries for Row$_{A}$. Moving Row$_{A}$ to its original location would displace Row$_{B}$. Row$_{B}$ is copied into the \emph{place-back} buffer, as shown by \encircle{3}.
\blue{The RIT updates the physical location of Row$_{B}$ in the real part as a place-back buffer, and the original location for Row$_{B}$ required for the next place-back operation can be obtained from the mirrored part of the RIT -- as shown by \encircle{4}.}

As shown by \encircle{5}, once Row$_{B}$ is moved to its original location, it displaces Row$_{C}$.  As shown by \encircle{6}, the RIT also invalidates the entries for Row$_{B}$. Thereafter, as shown by \encircle{7}, this process continues and Row$_{C}$ is copied into the \emph{place-back} buffer. 
\blue{As shown by \encircle{8}, this also updates the physical location of Row$_C$ in the real part as a place-back buffer.
The next stage moves Row$_{C}$ back to its original location and completes the lazy eviction process.
}
}

As shown in Figure~\ref{fig:srs-swapback-v2}, let us assume that the RIT contains mappings for Row$_{A}$, Row$_{B}$, and Row$_{C}$. If Row$_{A}$ is lazily evicted from the RIT, as shown by \encircle{~1 }, it will be first moved into the swap-buffers (already present in the original design of RRS~\cite{rrs}), as shown by \encircle{1.1}. Then, Row$_{B}$ is copied into the \emph{place-back} buffer. This is shown by \encircle{1.2}. Row$_A$ then moved to its original location, as shown by \encircle{1.3}. As the last step for the first place-back operation, the RIT invalidates the entries for Row$_A$ and updates the physical location of Row$_B$ in the \emph{real part} as the place-back buffer This is shown by \encircle{1.4}.

The next place-back operation moves Row$_B$ into its original location, as shown by \encircle{~2 }. Similar to the first place-back operation, as shown by \encircle{2.1}, it first moves the row (Row$_C$) in its original location into the swap buffer. The RIT invalidates the entries for Row$_C$, as shown by \encircle{2.2}. Now, Row$_B$ is moved into its original location, as shown by \encircle{2.3}. The RIT invalidates the entry for Row$_B$, as shown by \encircle{2.4}. Finally, Row$_C$ is migrated to its original location, and the lazy eviction process is completed. This is shown by \encircle{2.5}.

\ignore{
{\color{blue}
Will explain 

our HRT can track the number of accesses to physical row without knowing proprietary in-DRAM-mapping information. Thus, although SRRS is secure enough (the expected time for a successful attack is xx years even with TRH = 1600), SRRS can provide deterministic protection by co-designing with blacklisting method. Explain HRT's operation here with Figure for HRT operation.
}

\noindent \textbf{Row-Indirection Table (RIT):}
To avoid the security pitfalls in RRS, we introduce a novel RIT design.

{\color{blue} 
Will explain 
\begin{itemize}
    \item Our RIT can avoid reswap operations, so can avoid Juggernaut attack
    \item Our RIT has a lower performance overhead since (1) it only require swap operations not reswap operations, (2) and can amortize unswap overhead performing when the memory system is free
    \item RIT's operations, add Figure for RIT operation
\end{itemize}
}
}
\subsection{Security Analysis}\label{sec:security}
We quantitatively analyze the security of SRS against the Juggernaut attack pattern.
\begin{tcolorbox}
\textbf{Goal:} For a system with Secure Row-Swap (SRS), create a successful RH attack by causing any specific aggressor row ($Row_{aggr}$) to incur $\geq T_{RH}$ activations (ACTs).
\end{tcolorbox}

As illustrated in Section~\ref{sec::Juggernaut}, Juggernaut is composed of two parts. First, the attacker would attempt to bias any one aggressor row towards higher activation counts during $N$ attack rounds. However, since SRS employs the \emph{swap-only} row indirection, there are no additional latent activations on the \emph{original location} of the aggressor row in each round. Thus, the original location incurs only 1 latent activation (ACT) during the \emph{initial swap} operation of the aggressor row ($Row_{aggr}$). This is denoted by Equation~\ref{eqa:deterministic_accesses1}.
\begin{equation}~\label{eqa:deterministic_accesses1}
    ACT_{aggr}  = 2\times T_{S} 
\end{equation}

Since \emph{$Row_{aggr}$} already has received ACT$_{aggr}$ activations, the additional activations needed to cause this row to incur Row Hammer (ACT$_{left}$) are represented in Equation~\ref{eqa:additional1}. 
% We notice that the attacker now can throw nearly 2 $\times$ T$_{S}$ balls into the target bin, and this does not depend on $N_{attack}$.
\begin{equation}\label{eqa:additional1}
\begin{gathered}
    ACT_{left} = T_{RH} - ACT_{aggr} 
    % ACT_{left} = T_{RH} - (2\times T_{S} + 2)
\end{gathered}  
\end{equation}

Thereafter, the attacker uses the random-guess attack to pick random rows and activate them T$_S$ times. We explained this process in detail in Section~\ref{security}. The time for a successful attack can be obtained by plugging Equation~\ref{eqa:additional1} into Equation~\ref{eqa:k}.

\ignore{
Each \emph{swap} or \emph{reswap} operation presents an opportunity to the attacker(s) to choose a random bin (or Row$_{randN}$). If they pick the \emph{target bin} again, they could throw additional $T_{S}$ balls into the bin. We calculate the number of times the attacker has to land in the target bin to get $\geq T_{RH}$ balls in that bin -- essentially breaking RRS. This is determined by the number of \emph{swap} or \emph{reswap} operations of the target bin, denoted by `$k$', and is calculated using Equation~\ref{eqa:k1}.
\begin{equation}\label{eqa:k1}
\begin{gathered}
    k = \lceil \frac{ACT_{left}}{ T_S} \rceil
    % k = \lceil \frac{T_{RH} - (2\times T_{S}+2)}{ T_R} \rceil 
\end{gathered}  
\end{equation}

We then plug these numbers into Equation~\ref{eqa:Balls}, Equation~\ref{eqa:pkt}, Equation~\ref{eqa:ps}, Equation~\ref{Attackiteration}, and Equation~\ref{Attacktime} to develop the time for a successful targeted attack. 
}
\begin{figure}[h]
    \centering
    \vspace{-0.1in}
    \includegraphics[width=0.9\columnwidth]{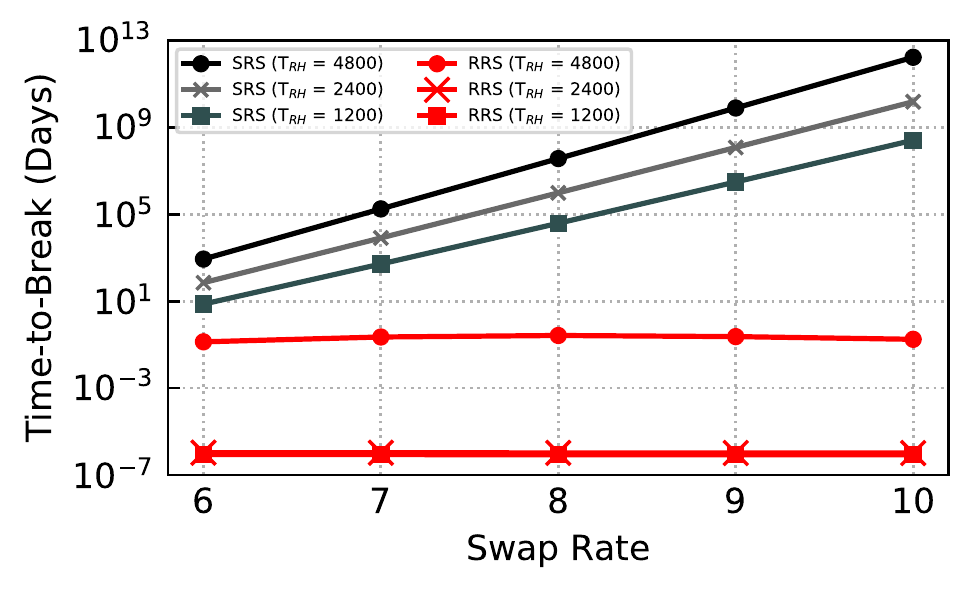}
    \vspace{-0.15in}
    \caption{Time-to-break SRS using the Juggernaut attack pattern. For T$_{RH}$ of 4800, even with a swap rate of 6, SRS has a time-to-break of $>2$ years while under continuous attack. In contrast, RRS can be broken in 4 hours.}
    \label{fig:srs_attacktime_swaprate_juggernaut}
    \vspace{-0.1in}
\end{figure}

Figure~\ref{fig:srs_attacktime_swaprate_juggernaut} shows the time-to-break SRS and RRS using Juggernaut as we increase the swap rate and vary T$_{RH}$ values. For a T$_{RH}$ of 4800, even with a swap rate of 6, SRS provides robust security for $>$2 years against the Juggernaut attack pattern. SRS is more robust at higher \emph{swap rates}. Unfortunately, even at increased swap rates, RRS is highly vulnerable to the Juggernaut attack pattern.

\subsection{Future-Proofing Security by Tracking Swap Counts}
To protect against any unknown future attack patterns, we future-proof SRS by adding a per-row swap-tracking counter. We reserve a small portion of the main memory to store these counters. Additionally, we also add a 19-bit on-chip register in the memory controller to count epochs. Similar to prior work, a refresh interval is divided into two epochs~\cite{park_graphene:_2020,hydra}. Each counter is composed of two parts. The first part stores an epoch-id. The second part stores the cumulative activation count when a swap occurs -- including any latent activations.  

\begin{figure}[t!]
    \centering
    \includegraphics[width=1\columnwidth]{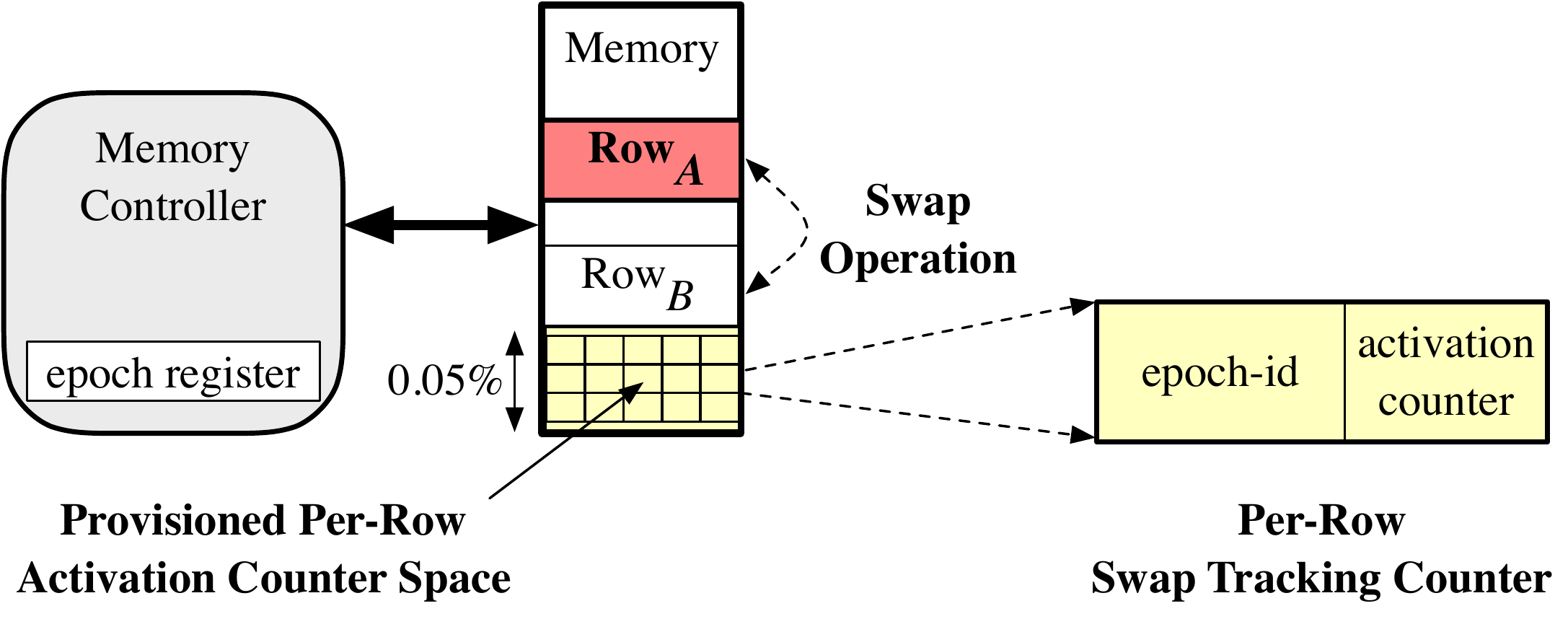}
    \caption{A memory system with per-row swap-tracking counter. The memory controller stores an epoch register. The main memory reserves 0.05\% of its space to store a per-row tracking counter. The respective counter for a row is read and updated before each swap operation.}
    \label{fig:futureproof}
    \vspace{-0.2in}
\end{figure}

Figure~\ref{fig:futureproof} shows this design. Let us assume that a counter with 19 bits of epoch-id and 13 bits of activation count. Therefore, it can count up to 8192 activations per row (including latent activations) per epoch. The respective counter for a row is read before a swap operation. If the on-chip epoch register is different from the 19-bit epoch-id, then it indicates a different epoch window. In this case, the activation counts for that row are reset. However, if the epoch-id and the on-chip epoch register have the same value, then activation counts are updated with T$_{S}$ activations along with any additional latent activation count. Once the on-chip epoch register shows all `1s', it immediately resets all the counters. This involves reading 64 counter rows every 2$^{19}$ epochs (each epoch is 32ms) -- incurring a latency of 41$\mu$s every 4.6 hours.

In terms of storage, we need only one 32-bit counter per DRAM row. Assuming we have 128K rows per bank, we would need to provision 512KB of space per bank. This represents 0.05\% of the total DRAM capacity. These 512KB of counters are stored across sixty-four 8KB DRAM rows accessed only during swap operations. To prevent any recursive look-ups, the counter-rows are tracked using dedicated per-bank on-chip activation counters (similar to prior work~\cite{hydra}).

\subsection{SRS: Performance and Scalability}
\begin{figure}[h!]
    \centering 
    \includegraphics[width=\columnwidth]{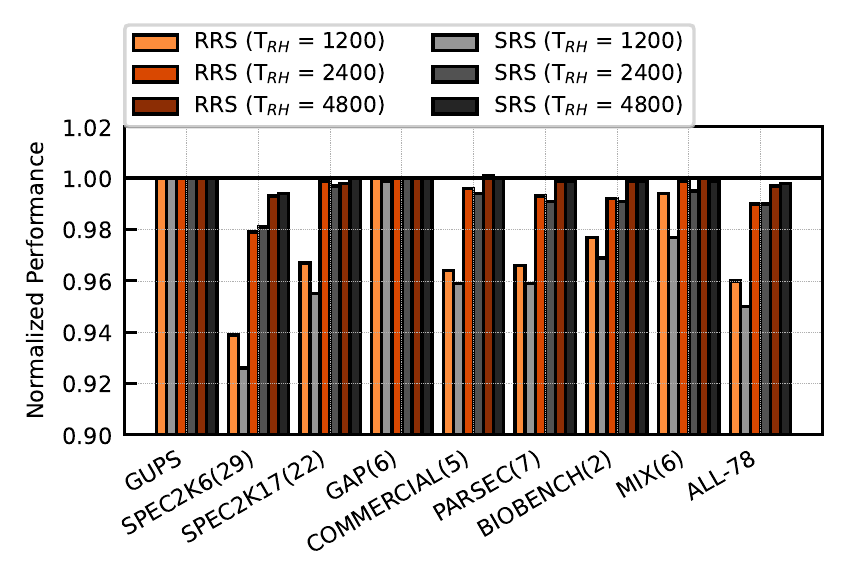}
    \vspace{-0.35in}
    \caption{The normalized performance of SRS and RRS compared to an not-secure baseline. Overall, SRS and RRS show similar slowdowns across different values of T$_{RH}$. The variation in performance occurs due to the sub-optimal schedules of the lazy eviction mechanism and place-back operations.}
    \label{fig:srsperf}
    \vspace{-0.2in}
\end{figure}

Figure~\ref{fig:srsperf} compares the performance of SRS with RRS. SRS shows a similar slowdown as RRS. This is because, while SRS prevents the Juggernaut attack, it still incurs the same memory bandwidth overheads as RRS. The memory bandwidth overheads are dictated by the swap rate. As the swap rate of SRS and RRS are the same, they do not scale well towards lower values of T$_{RH}$. SRS and RRS show a variation in performance occurs due to the sub-optimal schedules of the lazy eviction mechanism and place-back operations.

\section{Scalable and Secure Row-Swap}
\subsection{Overview and Intuition}
Scalable and Secure Row-Swap (Scale-SRS) aims to reduce the swap rate and mitigate the memory bandwidth overheads from swaps while providing years of security. To this end, Scale-SRS uses the observation that, even during an attack, the original locations of only a few aggressor rows receive multiple swaps. RRS and SRS increase the swap rate of the entire memory system \emph{only} to take care of these outlier rows. Instead of designing for the worse-case outlier rows, Scale-SRS designs for the common case. To this end, Scale-SRS detects the outlier rows and stores them in the Last Level Cache (LLC). Fortunately, even during an attack, there tend to be only a few outlier rows every few hours or days. Thus, the LLC observes a minor capacity loss only for one refresh interval that occurs every few hours or days (in the worst case).

\subsection{Improving Scalability by Reducing Swap Rates}
Even during an attack, there are only a few such locations that stand out as outliers. This is because, within a refresh window, there are only a finite number of activations ($ACT_{max} =$ 1.36 million) are possible. Assuming a T$_{S} =$ 1200, the attacker can only activate up to 1134 ($\frac{ACT_{max}}{T_{S}}$) rows T$_{S}$ times. Furthermore, if a T$_{RH}$ is 4800, then the attacker would need to land on the original location of any one of these rows 3 times. 

\begin{figure}[h!]
    \centering
    \includegraphics[width=0.8\columnwidth]{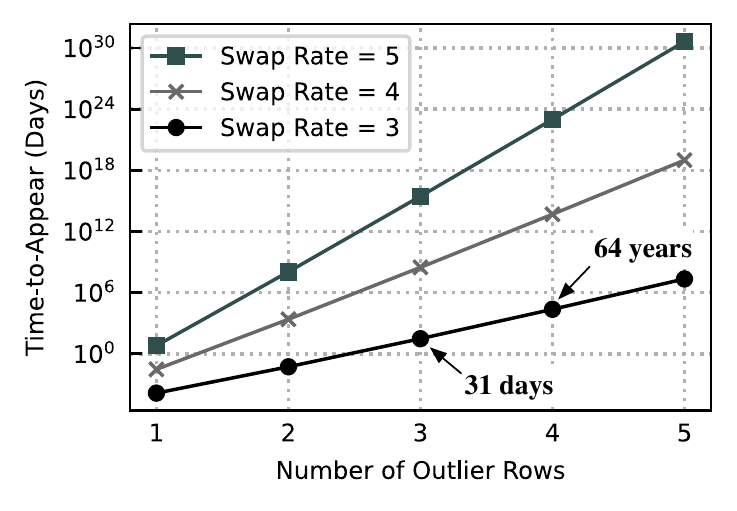}
    \vspace{-0.1in}
    \caption{The time-to-appear (in days) for outlier rows with varying swap rates for T$_{RH}$ of 4800. Even at a lower swap rate of 3, it takes at least \emph{64 years} for 4 outlier rows with $>$3 swaps to simultaneously appear within a bank. Additionally, only one 64ms refresh window every 31 days showcases 3 outlier rows -- thus, these outliers are very rare.}
    \label{fig:outlier}
    \vspace{-0.1in}
\end{figure}

Fortunately, the memory bank tends to have several rows -- say between 64K-128K rows. Even during an attack, only a small fraction of these rows (1134 rows) are swapped, and they have 64K-128K locations they could be swapped into. Thus, in most refresh intervals, the original location of any attacked rows would not have been chosen more than 3 times. The intervals wherein the row is chosen more than 3 times are outliers. These occur only every few hours or days. Figure~\ref{fig:outlier} shows the time to appear for these outlier rows with varying swap rates. For this analysis, we assumed a T$_{RH}$ of 4800.
\begin{figure*}
    \centering
    \includegraphics[width=\linewidth]{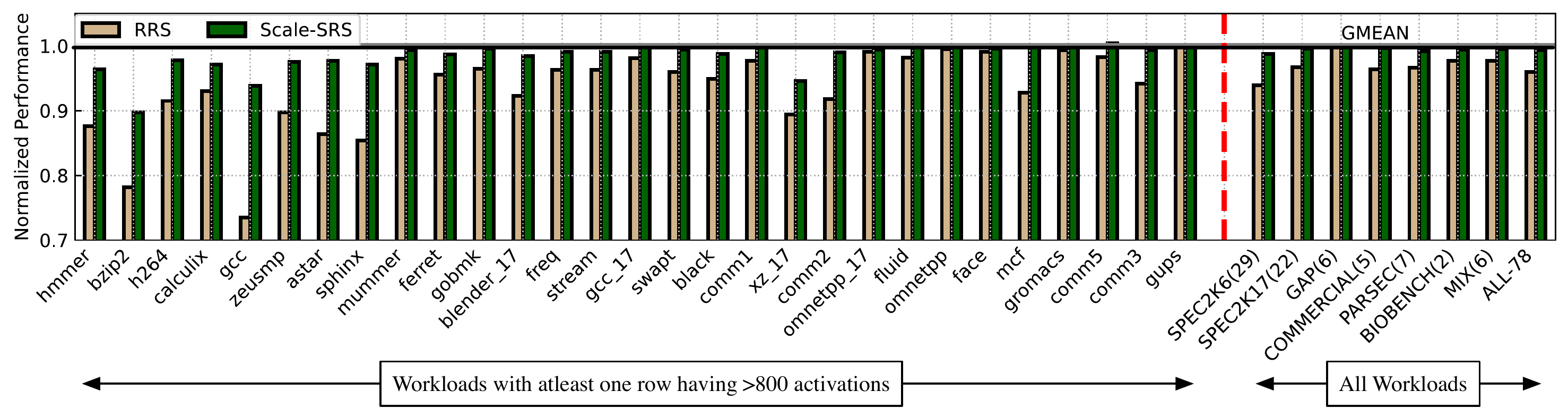}
    \caption{The normalized performance of Scalable and Secure Row Swap (Scale-SRS) and Randomized Row Swap (RRS) compared to a not-secure baseline at T$_{RH}$ of 1200. Scale-SRS and RRS incur an average slowdown of only 0.7\% and 4\% respectively, with several benchmarks in RRS incurring $>$10\% slowdown.}
    \label{fig:perf4800}
    \vspace{-0.15in}
\end{figure*}

Without loss of generality, this paper chooses a swap rate of 3. We observe that three rows (as shown in Figure~\ref{fig:outlier}) are chosen only three times in a 31-day window\footnote{The expected number of rows with `k' swaps for a DRAM bank that has `R' rows ($R_K$) is $R \times p_{k,T_S}$. The probability of having `M' rows with `k' swaps ($p_{M,k}$) can be calculated with the Poisson distribution as $\frac{{e}^{-R_{K}} \times R_K^{M}}{M!}$
}.
We use the per-row swap-tracking counters to identify such events. If any per-row swap-tracking counter value is $\geq 3 \times T_{S}$, we classify its respective row as an outlier and pin it within the LLC.

\subsection{Provisioning Space in the Last Level Cache}
Assuming a T$_{RH}$ of 4800, the LLC needs to be equipped to store a maximum of 3 DRAM rows in a single bank attack (occurring once every 31 days). As each row is 8KB and an adversary targets a single bank per channel (to maximize attack bandwidth), we may need up to 3$\times$8$\times$1$\times$2 = 48KB of space in the LLC. This accounts for only 0.05\% of 8MB LLC. 

We also analyze the multiple bank attack, as it might increase the capacity overhead in LLC. Assuming years of continuous attack, up to 3 outlier rows can appear in 11 banks per channel, which requires LLC to store 66 DRAM rows. For an 8MB LLC, this translates to a 6.5\% lower capacity.
However, as the multiple bank attack degrades the attack efficiency (as explained in Section~\ref{sec:juggeranut-attack-round}), \emph{this scenario now occurs only once every \textbf{2.6 years} and only lasts for \textbf{one refresh interval (64ms)}. Thus, on average, pinning rows in LLC has a negligible impact on performance.}

%(\textit{e.g.}, $\geq$15 years are required to cause 3 outlier rows per bank when targeting $\geq$12 banks)
% Assuming a T$_{RH}$ of 4800, the LLC needs to be equipped to store a maximum of 3 rows per bank (occurring once every 31 days).
% Fortunately, the attacker cannot target all 16 banks due to bank-to-bank activation delays, row migration latencies, etc. 
% We observe that it is most effective to target 11 banks per channel. As each row is 8KB and we have 11 banks with 2 channels, we may need up to 3$\times$8$\times$11$\times$2 = 528KB of space in the LLC. 
% This translates to roughly 66 DRAM rows. For an 8MB LLC, this translates to 6.5\% lower capacity. \emph{It is important to note that this scenario occurs only once every 31 days and only lasts for one refresh interval. Thus, on average, pinning rows in LLC has a negligible impact on performance}.

As the LLC employs its own address mappings into its sets, it cannot simply pin DRAM rows. It could be likely that these rows could map the same set and thereby conflict with each other. To prevent this, Scale-SRS employs a small buffer, called pin-buffer, in front of the LLC to indicate the pinned physical addresses and redirect them into their new set locations. For instance, we would need a 66-entry buffer that stores the addresses of 66 DRAM rows. For an 8KB row, each entry would be 35 bits long (48-bit physical address - 13-bits). 

Each pin-buffer entry points to a fixed set. For instance, the first entry would point to set 0. Assuming 64 Byte cache lines and an 8-way cache, we would need 16 contiguous sets to store this row. Thus, the second entry would now point to set 16, and so on. All accesses into the LLC flow through the pin-buffer, preventing any new cacheline from evicting these entries. These entries are cleared, and their respective rows are evicted once the refresh interval ends. In most 64ms refresh intervals, the pin-buffer does not contain any rows.
\ignore{
Other possible tables to add (Show the best configuration for each swap rate and T_RH)
\begin{tabular}{|c|c|c|c|}
T$_{RH}$    &	Swap Rate &		Outlier Rows Number  & Time to Appear    & LLC Overhead\\\hline\hline
\multirow{3}{*}{4800} &	3 &	3 & 30.7 days &  6.45\% \\
                      & 4 & 1	& 		3.13%
	5		1			3.13%
\multirow{3}{*}{2400} &	3		5			9.38%
	4		2			3.52%
	5		1			3.13%
\multirow{3}{*}{1200} &	3		7			9.38%
	4		3			4.69%
	5		1			3.13%
\end{tabular}

}

\ignore{
\begin{table}[h]
    \centering
    \caption{Time-to-Appear for Outlier Rows for Swap Rate $=$ 3}
    \resizebox{0.8\columnwidth}{!}{
    \begin{tabular}{|c|c|c|c|}
    \hline
    T$_{RH}$ & Outlier Rows Number & Time to Appear\\\hline\hline
     \multirow{3}{*}{4800}& 2  &   1.3 hours    \\
                          & 3  &   30.7 days    \\
                          & 4  &   64 years     \\\hline
     \multirow{3}{*}{2400}& 3  &   2.3 hours    \\
                          & 4  &   10.4 days    \\
                          & 5  &   3.9 years    \\\hline
     \multirow{3}{*}{1200}& 6  &   3.3 days\\
                          & 7  &   95 days\\
                          & 8  &   8.6 years \\\hline
    \end{tabular}}
    \label{tab:tta}
    \vspace{-0.1in}
\end{table}
}
\ignore{
{\color{blue}
Although Secure Row-Swap (SRS) can prevent the Juggernaut Attack-Pattern with almost half the RIT size of RRS, SRS still requires large SRAM overheads and incurs a significant performance overhead at lower values of RH thresholds ($T_{RH}$) and higher values of swap rates. For instance, at $T_{RH}$ of 1.2K and the swap rate of 8\footnote{Swap rate of 8 is chosen since it offers similar protection with $T_{RH}=4.8K$ and the swap rate of 6}, an overhead of 135KB of SRAM per bank is needed for the RIT in SRS, which is excessively large to implement in the memory controller. Moreover, SRS also causes a huge slowdown of $>$ ZZ\% even in benign applications, which makes SRS not scalable. To overcome this, we propose Scalable and Secure Row-Swap (Scale-SRS). In this section, we first describe our key observation to solve the scalability issue of SRS and provide an overview of Scale-SRS. We then determine the design parameters (swap rate and the number of reserved rows in the last-level cache (LLC)) required to minimize performance and storage overheads while preserving the security of SRS. 

\subsection{Intuition and Overview}
Will explain the key insight, overall explanation of Scale-SRS here.

\subsection{Analytical Model}
We use the Random Access attack pattern for our analysis since SRS is immune to Juggernaut and Juggernaut is effective to cause bit-flips only in a single row. Similar to Section~\ref{sec::Juggernaut} and~\ref{sec:security}, we use bins and balls analysis to decide the design parameters of Scale-SRS.

First, as shown in Section~\ref{random attack}, we can calculate the expected number of rows with k swaps (in 64ms) for a DRAM bank that has `N' rows as $N_{K} = N \times p_{k,T}$. Now, the probability of having `M' rows with k swaps ($p_{M,k}$) can be determined using the Poisson distribution. This is described by Equation~\ref{Poisson}.
\begin{equation}\label{Poisson}
    p_{M,k} = \frac{{e}^{-N_{K}} \times N_K^{M}}{M!}
\end{equation}

Finally, we obtain the expected number of attack iterations ($AT_{iter}$) to have `M' rows with k swaps as,
\begin{equation}
    AT_{iter} = \frac{1}{p_{M,k}}
\end{equation}

Since each attack iteration period is one refresh interval (64ms), the expected time required to gain `M' rows with k swaps (i.e., having `M' rows that experience bit-flips) is $AT_{time} = 64ms * AT_{iter}$. 
}

\subsection{Determining Required Number of Pinned Rows in the LLC}

Table~\ref{tab:attac_time_hammered_rows} shows the expected Attack Time ($AT_{time}$) to cause bit-flips on `M' rows at $T_{RH}=4.8K$. We show the Attack Time ($AT_{time}$) for different values of swap rates and numbers of hammered rows (M) in Table~\ref{tab:attac_time_hammered_rows}. At the same swap rate, as the number of hammered rows (M) increases, the time for a successful attack increases. For example, at the swap rate of 3, almost one month is required to cause bit-flips on 3 rows, but Attack Time ($AT_{time}$) increases to 64.1 years to incur 4 hammered rows in a refresh interval.  Thus, for each swap rate, we select the number of stored rows in the last-level cache (LLC) such that the system is protected at least 5 years of continuous attack (e.g., at the swap rate of 3, 3 rows space will be reserved in the LLC). 

As the swap rate increases, the required number of rows to be stored in the LLC reduces. However, there is a LLC overhead vs performance and the memory controller storage trade-off. This is because increasing the swap rate incurs performance and storage overheads because of frequent swap operations and larger RIT and tracker. Hence, we select the swap rate of each $T_{RH}$ of Scale-SRS to use $<$5\% of the LLC.

\begin{table}[h]
    \centering

    \caption{Required Attack Time ($AT_{time}$) to cause Row Hammer ($T_{RH}=4.8K$ activations) on multiple rows (M)}
    \begin{tabular}{|c|c|c|}
    \hline
     \textbf{Swap Rate (k)} & \textbf{Hammered Rows (M)} & \textbf{Attack Time ($AT_{time}$)}  \\\hline
     \multirow{5}{*}{3}     &   1       &   12.2 seconds\\
                            &   2       &   1.3 hours   \\
                            &   3       &   30.7 days   \\ 
                            &   4       &   64.1 years  \\
                            &   5       &   61135.3 years  \\\hline
     \multirow{5}{*}{4}     &   1       &   0.7 hours   \\
                            &   2       &   6.7 years  \\     
                            &   3       &   830,884 years  \\ 
                            &   4       &   1.3e11 years  \\ 
                            &   5       &   2.7e16 years \\\hline
     \multirow{5}{*}{5}     &   1       &   6.4 days   \\
                            &   2       &   304,657 years \\
                            &   3       &   7.8e12 years \\
                            &   4       &   2.7e20 years \\
                            &   5       &   1.2e28 years \\\hline
    \end{tabular}
    \label{tab:attac_time_hammered_rows}
\end{table}

We also analyze multi-bank attacks, where the attacker aims to cause bit-flips in multiple banks (maximum of 16). In this case, the theoretical number of rows to be reserved in the LLC would be a multiple of number of banks. However, as explained in RRS~\cite{rrs}, swap operations caused by attacking multiple banks decrease the attack time for each bank. We consider this aspect in our analysis.

\begin{table}[h]
    \centering

    \caption{TBD}
    \begin{tabular}{c|c|c|c}
    \hline
    \bf RH Threshold & \multirow{2}{*}{\bf Swap Rate} & \bf Total Reserved Rows & \bf LLC \\
     \textbf{($T_{RH}$)} & & \bf in the LLC & \bf Overhead  \\\hline
     \multirow{3}{*}{4.8K}& 3  &   66       &   6.5\% \\ 
                          & 4  &   32       &   3.1\% \\
                          & 5  &   32       &   3.1\%\\\hline
     \multirow{3}{*}{2.4K}& 3  &   96       &   9.4\% \\
                          & 4  &   36       &   3.5\% \\
                          & 5  &   32       &   3.1\% \\\hline
     \multirow{3}{*}{1.2K}& 3  &   96       &   9.4\% \\
                          & 4  &   48       &   4.7\% \\
                          & 5  &   32       &   3.1\% \\\hline
    \end{tabular}
    \label{tab:llc_overhead}
\end{table}

% JH: Not sure comparing tracker overhead is fair, since we don't introduce new tracker, and RRS can utilize Hydra Tracker also.
\begin{table*}[!bt]
  \centering
  \begin{small}
  \caption{TBD: Storage Overhead per Bank}
  \label{tab:storage_all_threshold}
  \begin{tabular}{|c||c|c|c||c|c|c||c|c|c|}
    \hline
\multirow{2}{*}{\bf Structure}  &   \multicolumn{3}{c||}{$T_{RH}=4.8K$}    &   \multicolumn{3}{c||}{\bf$T_{RH}=2.4K$}    &   \multicolumn{3}{c|}{\bf$T_{RH}=1.2K$} \\\cline{2-10}
	& \bf  RRS & \bf SRS & \bf Scale-SRS & \bf  RRS & \bf SRS & \bf Scale-SRS & \bf  RRS & \bf SRS & \bf Scale-SRS \\  \hline \hline
 
RIT             & 35 KB     & 18.8 KB   &   18.8KB   &   130KB  &   70KB  &   36.3KB  &   250KB   &   135KB    &   70KB\\ \hline
Tracker         & 6.9 KB    & TBD       &   TBD     &   23.8KB  &   TBD     &   TBD     &   42.5KB  &   TBD &   TBD\\ \hline
Swap-Buffers    & 1 KB      & 1 KB      &   1KB     &   1KB     &   1KB     &   1KB     &   1KB     &   1KB &   1KB\\ \hline 
Place-Back Buffers   & -    & 8 KB      &   8KB     &   -       &   8KB     &   8KB     &   -       &   8KB &   8KB\\ \hline 
Total           & 42.9 KB   & TBD       &   TBD     &   154.8KB  &   TBD     &   TBD     &   293.5KB &   TBD &   TBD\\ \hline \hline

LLC Overhead    &   -   &   -   &   3.1\%    &   -   &   -   &   3.5\%    &   -   &   -   &   4.7\%    \\\hline
  \end{tabular}
  \end{small}
\end{table*}

\subsection{Results: Performance Effectiveness of Scale-SRS}

\subsection{Results: Cost Effectiveness of Scale-SRS}

}
\ignore{
JH: I think we could reuse this part from previous submissions,
I also usually just paraphrase prior submissions if exists, but I'd like to learn how you do this efficiently
- Subsection 1: Simulation framework
    - Explanation about the simulator
    - Explanation about the system configuration, such as DRAM parameters, processor configurations, etc (Include a table)
- Subsection 2: Workload
    - What Workload we used
    - How many instructions are performed 
}

\section{Evaluation Methodology}
\noindent\textbf{Simulation Framework:}  
We use a detailed memory system simulator USIMM~\cite{usimm,msc}, which is modified to enforce the DDR4 protocol. The Misra-Gries tracker and the RIT are modeled as a Collision Avoidance Table (CAT) structure~\cite{rrs} within the memory controller. We report the performance and other related metrics from the USIMM memory model. %and power models~\cite{micron_ddr4}. We use Cacti 6.0~\cite{cacti} with 32~nm technology to report the SRAM power of on-chip structures.

Table~\ref{table:system_config} shows the baseline system configuration. We use a DRAM configuration with 16 banks per rank and 1 rank per channel (similar to the prior work~\cite{rrs}) and 2 channels. Each bank has 128K rows of 8KB each and 1.36 million activations possible per bank in the 64ms refresh interval. To emphasize the scalability of Scale-SRS, we evaluate against a T$_{RH}$ of 1200 activations. We also perform sensitivity studies for T$_{RH}$ values of 512, 2400, and 4800 activations.

\begin{table}[h!]
\vspace{-0.1in}
\begin{center}
\begin{small}
\caption{Baseline System Configuration}{
\resizebox{0.8\columnwidth}{!}{
\begin{tabular}{|c|c|}
\hline
  Cores (OoO)           & 8        \\
  Processor clock speed        & 3.2GHz    \\
  ROB size           & 192       \\
  Fetch and Retire width & 4         \\ \hline
  Last Level Cache (Shared)    & 8MB, 16-Way, 64B lines \\ \hline
  Memory size                  & 32 GB -- DDR4 \\
  Memory bus speed             & 1.6 GHz (3.2GHz DDR) \\
  T$_{RCD}$-T$_{RP}$-T$_{CAS}$ & 14-14-14 ns\\
  T$_{RC}$, T$_{RFC}$, T$_{REFI}$      & 45ns, 350 ns, 7.8$\mu$s \\
 % T$_{REFI}$                   & 7.8$\mu$s \\
  Banks x Ranks x Channels     & 16 x 1 x 2 \\
  Rows per bank                & 128K \\ 
  Size of row                  & 8KB  \\ \hline

\end{tabular}}
\label{table:system_config}
}
\end{small}
\end{center}
\end{table}

\noindent\textbf{Workloads:} We evaluate Scale-SRS across SPEC2006~\cite{SPEC2006}, SPEC2017~\cite{SPEC2017}, GAP~\cite{GAP}, BIOBENCH~\cite{BIOBENCH}, PARSEC~\cite{PARSEC}, and COMMERCIAL~\cite{usimm} benchmarks. We use Intel Pintool~\cite{luk2005pin} to extract the SPEC2006, SPEC2017, and GAP benchmarks for representative regions. The COMMERCIAL, BIOBENCH, and PARSEC benchmark traces are obtained from the USIMM distribution. We executed each benchmark for 1 Billion instructions per core. We also create 6 mixed workloads by randomly combining benchmarks. We execute the workloads in rate mode and continue simulating the individual benchmarks until all cores complete 1 billion instructions each. For conciseness, we show detailed results only for workloads that encounter at least one row with 800+ activations within a 64ms time refresh window and report averages for all 78 workloads.

\ignore{
\begin{table}[!htb]
  \centering

  \caption{Workloads Characteristics (with Rows ACT-800+)}
  \label{table:activations}
 \resizebox{1\columnwidth}{!}{
  \begin{tabular}{|c||c|c|c|}
    \hline
Workload	&	Footprint (GB)	&	MPKI	&	Rows ACT-800+ \\ \hline \hline
hmmer	&	0.01	&	0.84	&	1675	\\
bzip2	&	2.41	&	5.57	&	1150	\\
h264	&	0.05	&	0.52	&	1136	\\
calculix	&	0.16	&	1.12	&	932	\\
gcc	&	0.09	&	4.42	&	818	\\
zeusmp	&	0.55	&	2.00	&	405	\\
astar	&	0.04	&	1.04	&	352	\\
sphinx	&	0.13	&	12.90	&	242	\\
mummer	&	2.17	&	19.13	&	192	\\
ferret	&	0.79	&	5.67	&	132	\\
gobmk	&	0.2	&	1.17	&	79	\\
blender\_17	&	0.24	&	1.53	&	53	\\
freq	&	0.59	&	2.89	&	44	\\
stream	&	0.63	&	3.48	&	41	\\
gcc\_17	&	0.36	&	0.55	&	38	\\
swapt	&	0.76	&	3.52	&	37	\\
black	&	0.55	&	3.08	&	37	\\
comm1	&	1.55	&	5.93	&	19	\\
xz\_17	&	0.64	&	5.12	&	12	\\
comm2	&	3.37	&	6.14	&	8	\\
omnetpp\_17	&	1.55	&	9.81	&	7	\\
fluid	&	0.99	&	2.70	&	7	\\
omnetpp	&	1.1	&	17.24	&	5	\\
face	&	1.1	&	7.18	&	3	\\
mcf	&	7.71	&	107.81	&	2	\\
gromacs	&	0.06	&	0.58	&	1	\\
comm5	&	0.67	&	1.48	&	1	\\
comm3	&	1.77	&	2.84	&	1	\\ \hline
  \end{tabular}}
\end{table}
}

\begin{comment}
\begin {table}[h]
\begin{center} 
\caption{Workload Mixes}{
\resizebox{1\columnwidth}{!}{
\begin{tabular}{|c||c|}
\hline
mix1 & mummer, lbm, comm1, black, nab\_17, cc\_t, hmmer, leela\_17\\
mix2 & tigr, povray, comm2, ferret, xz\_17, cc\_w, wrf, gcc\_17\\
mix3 & h264, mcf, comm3, face, namd\_17, bc\_w, sphinx, cactuBSSN\_17\\
mix4 & comm4, gcc, fluid, bc\_t, mummer, omnetpp, fotonik3d\_17, roms\_17\\
mix5 & libquantum, comm5, freq, pr\_t, comm3, tigr, x264\_17, cam4\_17 \\
mix6 & milc, gups, stream, xalancbmk, parest\_17, pr\_w, leslie3d, swapt\\
\hline
\end{tabular}}
}
\label{table:mix_workload}
\end{center}
\end{table}
\end{comment}

\section{Results and Analysis}
\subsection{Performance}
Figure~\ref{fig:perf4800} shows the normalized performance of Scale-SRS and RRS with respect to a baseline that does not employ RH mitigation. To emphasize the scalability of Scale-SRS, we use an aggressively low T$_{RH}$ of 1200. Workloads such as \texttt{hmmer}, \texttt{bzip2}, \texttt{gcc}, \texttt{zeusmp}, \texttt{astar}, \texttt{sphinx}, and \texttt{xz\_17} have greater than 10\% slowdown while employing RRS. In the worst case, \texttt{gcc} has a 26.5\% slowdown due to frequent swaps in RRS. On average, across 78 workloads, Scale-SRS has a slowdown of only 0.7\%, whereas RRS has a slowdown of 4\%.

\subsection{Sensitivity to Varying RH Thresholds}
\begin{figure}[h!]
    \vspace{-0.15in}
    \centering
    \includegraphics[width=\columnwidth]{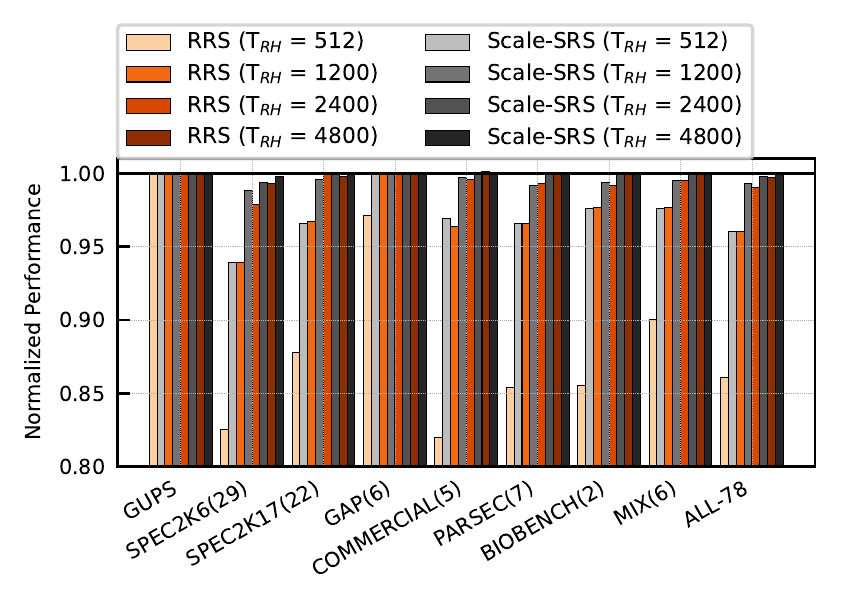}
    \vspace{-0.35in}
    \caption{The normalized performance of SRS and RRS as the value of T$_{RH}$ varies from 4800 to 512. Even at a T$_{RH}$ of 512, Scale-SRS shows an average slowdown of only 4\%, whereas RRS shows an average slowdown of 14\%.}
    \label{fig:thresholds}
\end{figure}

Figure~\ref{fig:thresholds} shows the performance sensitivity of Scale-SRS and RRS as T$_{RH}$ varies from 4800 to 512. Even when T$_{RH}$ drops, Scale-SRS minimizes its performance overhead since it employs a relatively lower swap rate. On the contrary, RRS incurs higher performance overhead as RRS caters to the outlier rows, which makes it swaps (and unswaps) rows at a relatively higher rate.
% On the contrary, as RRS caters to the outlier rows, it swaps (and unswaps) rows at a relatively higher rate. 
Even at a T$_{RH}$ of 512, Scale-SRS shows an average slowdown of only 4\%, whereas RRS shows an average slowdown of 14\%.

\subsection{Impact of Aggressor Row Tracker}
\begin{figure}[h!]
    \centering
    \includegraphics[width=\columnwidth]{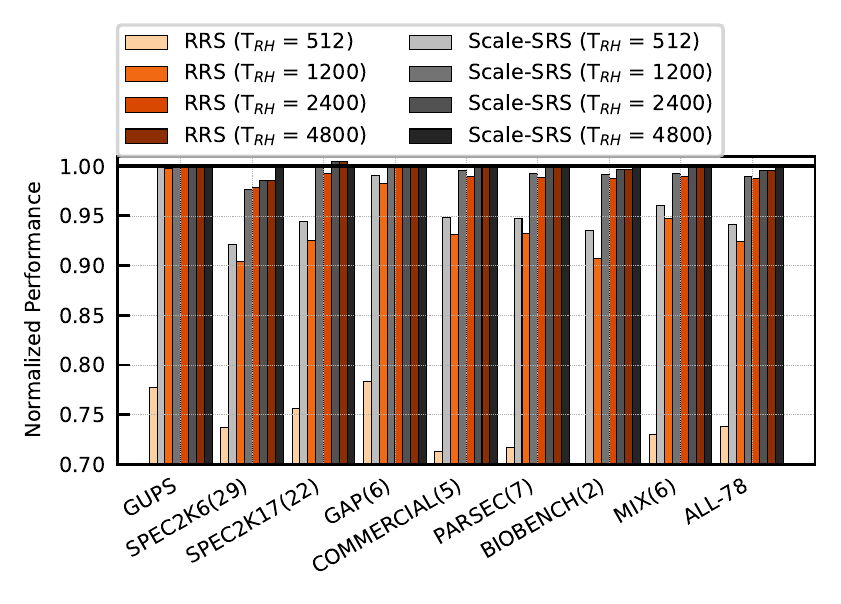}
    \caption{The normalized performance of Scale-SRS and RRS while using the Hydra tracker. Even at a T$_{RH}$ of 512, Scale-SRS with Hydra has an average slowdown of only 5.9\%, whereas RRS has an average slowdown of 26.8\%.}
    \label{fig:hydra}
\end{figure}

Figure~\ref{fig:hydra} shows the performance sensitivity of Scale-SRS and RRS if they use the Hydra tracker instead of the Misra-Gries Tracker. We vary T$_{RH}$ from 4800 to 512. Even at a T$_{RH}$ of 512, Scale-SRS with Hydra has an average slowdown of only 5.9\%, whereas RRS has an average slowdown of 26.8\%. Hydra stores its activation counters in the memory. Thus, despite using a counter cache, RRS with Hydra tends to access the memory frequently at lower T$_{RH}$ values. 

\subsection{Storage Analysis}
Table~\ref{tab:storage_all_threshold} shows the required SRAM-based on-chip storage for RRS and compares that to Scale-SRS. A key difference between RRS and Scale-SRS is the reduced swap rate of 3. This enables Scale-SRS to reduce the size of the RIT. 

Scale-SRS requires one additional 8KB place-back buffer per bank. Additionally, it also uses a 19-bit epoch register and a pin-buffer. The size of the pin-buffer depends on the number of outlier rows -- which is determined by T$_{RH}$. The LLC overhead from pinning rows occurs only once every few thousand 64ms refresh intervals. Thus, it has a negligible impact on performance and is not shown in Table~\ref{tab:storage_all_threshold}. Overall, Scale-SRS has about 3.3$\times$ less storage overhead compared to RRS at a T$_{RH}$ of 1200.

\begin{table}[!h]
  \vspace{-0.1in}
  \centering
  \begin{small}
  \caption{Storage Overhead per Bank}
  \label{tab:storage_all_threshold}
   \resizebox{\columnwidth}{!}{
  \begin{tabular}{|c|c|c|c|c|c|c|}
    \hline
\multirow{2}{*}{\bf Structure}  &   \multicolumn{2}{c|}{$T_{RH}=4800$}    &   \multicolumn{2}{c|}{$T_{RH}=2400$}    &   \multicolumn{2}{c|}{$T_{RH}=1200$} \\%\cline{2-8}
	& \bf  RRS & \bf Scale-SRS & \bf  RRS & \bf Scale-SRS & \bf  RRS & \bf Scale-SRS \\  \hline \hline
 
RIT             & 35 KB    &   9.4KB   &   130KB  &   35KB  &   250KB &   67.5KB\\ \hline
Swap-Buffer    & 1 KB      & 1 KB      &   1KB     &   1KB     &   1KB     &   1KB \\ \hline 
Place-Back   & \multirow{2}{*}{-}& \multirow{2}{*}{8KB}& \multirow{2}{*}{-} & \multirow{2}{*}{8KB} &  \multirow{2}{*}{-}  &   \multirow{2}{*}{8KB} \\
Buffer   & & & & & &     \\ \hline
Epoch & \multirow{2}{*}{-}& \multirow{2}{*}{19 bits}& \multirow{2}{*}{-} & \multirow{2}{*}{19 bits} &  \multirow{2}{*}{-}  &   \multirow{2}{*}{19 bits} \\
Register   &  &  &  &  &  &  \\ \hline
Pin Buffer & - & 289 bytes &  - & 420 bytes  & - &  420 bytes \\ \hline\hline
Total  & 36 KB  &   18.7KB     &   131KB  &   44.4KB & 251KB & 76.9KB \\ \hline
  \end{tabular}}
  \end{small}
\vspace{-0.1in}
\end{table}

\ignore{
 \begin{table}[!htb]
   \centering
   \begin{small}
   \caption{Storage Overhead Per Bank (T$_{RH}$ = 1200)}   \label{tab:storage}
   \begin{tabular}{|c||c|c|}
     \hline
 \bf Structure	& \bf  RRS & \bf Scale-SRS \\  \hline \hline
 
 RIT             & 35 KB     & 17.5 KB  \\ \hline
 Swap-Buffers    & 1 KB      & 1 KB     \\ \hline 
 Place-Back Buffers   & -    & 8.01 KB  \\ \hline 
 Epoch Register & - & 19 bits \\ \hline
 Pin-Buffer & &
 Total           & 42.9 KB   & 33.4 KB  \\ \hline
   \end{tabular}
   \end{small}
 \end{table}
}
\subsection{Power Analysis}
Scale-SRS incurs power overheads from extra operations such as row swaps and accesses to on-chip structures. Table~\ref{table:power} shows the power consumed by DRAM (obtained from USIMM~\cite{usimm}) and the SRAM structures (obtained using Cactii~\cite{cacti} in the 32~nm technology) in Scale-SRS and RRS. Compared to RRS, due to smaller-sized SRAM structures, Scale-SRS incurs 23\% lower on-chip power. Scale-SRS also reduces the DRAM power as it reduces the effective swap rate.

\begin{table}[!h]
  \centering
  \begin{small}
  \caption{Extra Power Consumption Per Channel (T$_{RH}$ = 4800)}
  \vspace{-0.1 in}
  \label{table:power}
 \resizebox{1\columnwidth}{!}{
  \begin{tabular}{|c|c|c|c||c|}
    \hline
 \multirow{2}{*}{\bf Type of Power Overhead}	&\multirow{2}{*}{\bf RRS} & \bf Scale  \\ 
    & & \bf SRS  \\ \hline \hline
DRAM Power Overhead (Row-Swap)	        &	0.5\%   & 0.2\%     \\ \hline
SRAM Power Overhead	&	903 mW  & 703 mW  \\ \hline
  \end{tabular}
  }
  \end{small}
  \vspace{-0.1in}
\end{table}

\section{Discussion}
\noindent\textbf{1. Internal Chip Address versus Physical Address}:\\ 
We have demonstrated Scale-SRS and RRS using physical addresses supplied by the OS. However, it is possible that the chip rows are larger. In such scenarios, the memory controller can use the chip row addresses for the RIT and swap these rows. While this requires knowledge of the internals of DRAM, this does not change our technique or the security analysis.
\vspace{0.05in}

\noindent\textbf{2. Implementing Scale-SRS Near-Memory or In-Memory}:\\ 
While we have demonstrated Scale-SRS on the CPU-based memory controller, it does not prevent us from implementing this as near-memory or in-memory (within DRAM chips~\cite{lisa-indram,ertr}). This can help new interfaces such as CXL~\cite{cxl}.
\vspace{0.05in}

\noindent\textbf{3. Juggernaut Attack with Open-Page Policy}:\\
Using an open-page policy~\cite{kaseridis2011minimalist} for the memory controller could reduce the attack potency of Juggernaut. This is because keeping the page open can reduce the number of row activations and thereby decrease the maximum number of possible attack rounds. For instance, using open page policy at a T$_{RH}$ of 4800 and a swap rate of 6, the time-to-break RRS using Juggernaut increases from 4 hours to 10 days. However, the advantages of using open page policy \emph{disappear as T$_{RH}$ decreases}. At lower T$_{RH}$ values, Juggernaut is powerful \emph{regardless of page policies}. For example, if T$_{RH} \leq 3300$, Juggernaut can break RRS in under 1 day, even with the swap rate of 10. Thus, developing a new protection method against Juggernaut, such as our Scale-SRS, is essential to enable the adoption of randomized-based defense in the future DRAM generations (with lower T$_{RH}$).
\vspace{0.05in}

\noindent\textbf{4. Possible Storage Overhead Reduction of Scale-SRS:}\\
Although Scale-SRS has much less SRAM-based storage overhead than RRS, there is still room for storage overhead reduction. One way is to add a bit to every RIT entry to distinguish between the original and the reverse mapping. This would prevent the need for a mirrored part of the RIT and can reduce its storage overhead by almost 2$\times$.
\vspace{0.05in}

\noindent\textbf{5. Juggernaut and Scale-SRS in Future DRAM Generations:}\\
The T$_{RH}$ value will highly likely drop further in future DRAM generations, making them more vulnerable to RH-based attacks such as Juggernaut and half-double. Thus, future DRAM generations would involve more features to mitigate Row Hammer. For instance, recently introduced DDR5 devices perform refresh operations 2$\times$ more frequently than DDR4~\cite{refreshpause1,refreshpause2}. However, even in DDR5 devices, Juggernaut can break RRS in under 1 day regardless of the swap rate if T$_{RH} \leq 3100$. This demonstrates the potency of the Juggernaut attack even for future DRAM generations. This also highlights the necessity of new protection methods such as Scale-SRS. Furthermore, Scale-SRS has better scalability (\textit{i.e.}, better performance and less storage overhead) than RRS at lower T$_{RH}$ values. This enables Scale-SRS to be commercially viable as a defense line against RH attacks (known and unknown) for present and future DRAM generations.
%\newpage
\ignore{
Flow JH
We have already described and analyzed the most closely related RowHammer mitigation mechanisms in {Add section here}. In this section, we briefly discuss other related works.
Subsection 1: Probabilistic Methods for RowHammer Mitigation
    - Explain what probabilistic methods are: mitigate RH by carrying out additional refreshes with a certain probability
    - Explain each method and their limitations briefly
        Common things: All of probabilistic methods can be implemented with simple HW logic, but 
        (1) they cannot provide guaranteed protection due to their probabilistic nature, and (2) requires proprietary in Dram mapping.
        PARA (ISCA'14): Refreshes a victim row (neighboring row) with a predetermined probability, p, for every access. Can provide a more robust protection with a higher probability p, but it incurs severe performance overhead. 
        PRoHIT (DAC'17): Utilizes a probabilistic history tables to track potential victim rows more efficiently.      However, it is 1) not secure, especially when the # of aggressor rows > table size, and 2) incurs significant performance overhead even in benign applications.
        MRLoc (DAC'19): Utilizes a circular queue as a history table to distinguish victim rows more efficiently, but it is not secure, especially when the # of aggressor rows > queue size since it cannot track the memory access pattern. 
Subsection 2: SW-based RH mitigation (Might be able to reuse the content in the RRS paper)
    - Practical since it can be also adopted in existing systems.
    - However, (1) often requires knowledge of in dram mapping, (2) incurs a high peformance overhead, (3) only effective for a certain type of attack, (4) often requires a lot of modifications to system software.
Subsection 3: RH-based attacks
    - Introduce several examples of RH-based attacks, like Black Hat (2015), RowHammer js (2016), TRRespass (SP'20), Google's Half-double (2021), Blacksmith (SP'22).
    - Claims that our idea can protect all of these attacks.
}

\section{Related Work}
\label{sec:related}
% In this section, we discuss the closely related works.
\subsection{Aggressor-Focused Mitigation}
We have already described and analyzed the most closely related state-of-the-art aggressor-focused mitigation, Randomized Row-Swap (RRS), in Section~\ref{rrs-mitigation}. Besides RRS, BlockHammer (BH)~\cite{yauglikcci2021blockhammer} is another aggressor-focused mitigation. BH exploits dual counting bloom filters to track potential aggressor rows and uses a throttling-based approach for such rows. Unfortunately, BH is vulnerable to denial-of-service (DoS) attacks. For instance, at a T$_{RH}$ of 4800, memory requests would be delayed by approximately 20$\mu$s per activation. BH also requires complex memory scheduling policies. In comparison to BH, Scale-SRS is more efficient and has no DoS concerns. A recent work, AQUA~\cite{saxena2022aqua}, improves the performance and storage overhead of RRS by exploiting isolation instead of randomization. Specifically, AQUA reserves a dedicated region of DRAM as the quarantine region and migrates the aggressor rows into the quarantine region when the migration threshold is reached. As compared to AQUA, Scale-SRS does not need a dedicated quarantine region and relies on randomized row movement.

\subsection{Victim-Focused Mitigation}
Victim-focused mitigation (VFM) prevents RH by performing targeted refreshes on victim rows. This can be done either probabilistically (PRA~\cite{kim2014architectural}, PARA~\cite{kim2014flipping}, PRoHIT~\cite{PROHIT}, MRLoC~\cite{MRLOC}, HammerFilter~\cite{hammerfilter}) or by tracking accesses to particular rows (CRA~\cite{kim2014architectural}, CBT~\cite{CBT}, TWiCe~\cite{lee2019twice}, Graphene~\cite{park_graphene:_2020}, Hydra~\cite{hydra}).
While it is effective to prevent classic RH attacks that target victims that are immediate neighbors, they are susceptible to attack patterns, such as the \textit{half-double} attack~\cite{half-double, half-double2}, that target distant neighbors. One way that VFM may adapt to defend against \textit{half-double} is to account for neighbor refreshes in the activation counts of the tracker. However, this requires VFM to know the proprietary internal DRAM row mappings and accurate theoretical modeling of the \textit{half-double} and blast-radius effects. To the best of our knowledge, these effects are not yet fully known.

Mithril~\cite{mithril} and ProTRR~\cite{marazzi2022protrr} suggest using the newly introduced Refresh Management (RFM)-based RH mitigations. These solutions are implemented inside DRAM chips and coordinate with the memory controller using the RFM interface. This approach solves the limitations of prior VFM methods (such as requiring proprietary internal DRAM row mappings or an additional interface to communicate with the memory controller). ProTRR also shows how to prevent the \emph{half-double} attack. However, as T$_{RH}$ becomes lower and blast-radius increases due to DRAM technology scaling, implementing these methods inside DRAM chips tends to become infeasible due to their high storage overhead.

\ignore{
\red{However, most VFM methods that reside in the memory controller, such as Graphene and PARA, require the proprietary internal DRAM mappings to issue refreshes to victim rows. Recently, Mithril~\cite{mithril} introduced 

the recent \textit{half-double} attack shows that an attacker could exploit the mitigative actions of VFM to induce bit flips in distant neighbors. One way that VFM could adapt to defend against this kind of attack is by including the targeted refreshes in the activation counts of the tracker.}

\begin{table}[!htb]
  \centering
 % \vspace{-0.15 in}
  \begin{small}
  \caption{Comparison of RRS with Victim-Focused Mitigation}
  \vspace{-0.1 in}
  \label{table:VictimFocused}
  \setlength{\tabcolsep}{3pt}
  \begin{tabular}{ccc}
    \hline
    \textbf{Attribute} & \textbf{Victim-Focused} & \textbf{RRS}\\ \hline \hline
    Slowdown  &    $<$0.1\% & 0.4\% \\ \hline
 %   Storage per Bank  &  3KB (680-entry CAM) & 45KB \\ \hline
    Mitigates Classic Row Hammer & \multirow{2}{*}{\cmark} & \multirow{ 2}{*}{\cmark} \\
    (Neighboring Row Bit-Flips) & & \\ \hline
    Mitigates Complex Patterns & \multirow{2}{*}{\xmark} & \multirow{ 2}{*}{\cmark} \\ 
     (Far Aggressors of Half-Double~\cite{half-double}) & & \\ \hline
    Works Without Knowing & \multirow{2}{*}{\xmark} & \multirow{ 2}{*}{\cmark} \\ 
    DRAM Mapping           & &  \\ \hline
  \end{tabular}
  \end{small}
  \vspace{-0.1 in}
\end{table}
}

\subsection{ECC-Based Defenses}
ECC memories can correct a small number of bit-flips~\cite{avatar,xed,citadel1,citadel2,morphecc}. Such an approach can be used to correct the bit-flips from RH. However, ECCploit~\cite{cojocar2019exploiting} shows that an attacker can still cause RH by overcoming ECC protection. Synergy~\cite{saileshwar2018synergy} and SafeGuard~\cite{ali2022safeguard} provide integrity protection and can detect RH without recovering corrupted data.

\subsection{Software-Based Defenses}
Software-based defenses often require information about DRAM properties that may be proprietary or not readily accessible to software~\cite{aweke2016anvil, bock2019rip, konoth2018zebram, van2018guardion}. Additionally, these solutions often incur severe performance overheads, demand intrusive modifications to system software, and only tend to be effective for certain types of attacks.

For example, ANVIL~\cite{aweke2016anvil} employs CPU performance counters to identify RH attacks and perform refreshes to the immediate victim rows. GuardION~\cite{van2018guardion} prevents RH attacks by putting a guard row between data of different security domains. In ZemRAM~\cite{konoth2018zebram} and RIP-RH~\cite{bock2019rip}, isolation is provided by locating the kernel space and user space(s) in isolated parts of DRAM. Unfortunately, these solutions require proprietary internal DRAM mappings information. Other solutions, such as CATT~\cite{brasser2017can}, which carries out profiling of cells and blacklists pages that contain vulnerable cells to RH, can cause significant loss of memory capacity at lower T$_{RH}$. 

%A prior work, Monotonic Pointers~\cite{wu2019protecting}, exploits the fact that cells are prone to Row Hammer in one direction. However, this approach only protects PTEs and is vulnerable to RH-based inter-process attacks that break process isolation. %It uses this information to induce bit-flips in page-table-entries (PTEs) predictable. 

%Software-based defenses maybe not be capable of fully mitigating the root cause of the RH vulnerability in DRAM, leaving them vulnerable to newer attacks. For instance, a recently proposed \textit{half-double} attack~\cite{half-double,half-double2} could break the refresh of immediate victims of ANVIL or a single guard-row of GuardIon. Despite isolation between user and kernel spaces in RIP-RH and ZebRAM, user-space code could continuously cause bit-flips in kernel memory with attacks like PTHammer~\cite{zhang_pthammer:_2020}, which hammer kernel memory by initiating frequent page-table walks from user space.

\section{Conclusion}
As DRAM-based systems are becoming increasingly susceptible to Row Hammer (RH) attacks, a recent work called Randomized Row-Swap (RRS) proposed proactively swapping aggressor rows to break spatial correlations with victim rows. Our paper shows that RRS neither secure nor scalable. We propose Juggernaut that breaks RRS in under \emph{1 day} regardless of the swap rate. Juggernaut uses latent activations in RRS to make a row vulnerable to RH. To overcome this, we propose the Scalable and Secure Row-Swap (Scale-SRS). Scale-SRS avoids latent activations and prevents Juggernaut. 
It also enables scalable RH mitigation by allowing the use of a much lower swap rate than RRS. Overall, even at an RH threshold of 1200, Scale-SRS has a 0.7\% slowdown while requiring 3.3$\times$ less on-chip storage compared to RRS, which has a 4\% slowdown.

\section*{Acknowledgements}
This project is a part of the Systems and Architecture Laboratory (\textit{STAR Lab}) at the University of British Columbia (UBC). We express our thanks to the entire team of the Advanced Research Computing (ARC) Center at UBC~\cite{sockeye}. We also thank the anonymous reviewers and Moinuddin Qureshi for their invaluable feedback and comments. This work was partially supported by the Natural Sciences and Engineering Research Council of Canada (NSERC) [funding reference number RGPIN-2019-05059] and a Gift from Meta Inc. The views and conclusions contained herein are those of the authors and should not be interpreted as necessarily representing the official policies or endorsements, either expressed or implied, of NSERC, the Canadian Government, Meta Inc., NVIDIA, Georgia Institute of Technology, or UBC.

%\newpage
\appendix
\hl{
\section{Artifact Appendix}

%%%%%%%%%%%%%%%%%%%%%%%%%%%%%%%%%%%%%%%%%%%%%%%%%%%%%%%%%%%%%%%%%%%%%
\subsection{Abstract}
This artifact covers two aspects of the results from the paper: (1) Security analysis of our Juggernaut attack against Randomized Row-Swap (RRS) and (2) Performance analysis of our Scalable and Secure Row-Swap (Scale-SRS) and RRS.

For the security analysis, a Bins and Buckets model of the Juggernaut attack is provided as a C++ program. Our program is based on event-driven Monte Carlo simulations for faster simulations. We provide scripts to compile our simulators and to recreate the results shown in \cref{fig:attacktime_vs_round}.
%since naive Monte Carlo Simulations could spend a lot of time, like a few days or a week, for the simulations

For the performance analysis, we provide the C code for the implementation of Scale-SRS and RRS, which is encapsulated within the USIMM~\cite{usimm} memory system simulator.
The Scale-SRS and RRS structures and operations are implemented within the memory controller module in our artifact.
We provide scripts to compile our simulator and run the baseline, Scale-SRS, and RRS for all the workloads and plot the results in \cref{fig:perf4800}.

\subsection{Artifact Check-List}
\noindent \subsubsection{Security Evaluations}
{\small
\begin{itemize}
  \item {\bf Algorithm: } Implementation of event-driven Monte Carlo Simulations of the Juggernaut attack in C++. 
  \item {\bf Compilation: } Tested with g++ (versions 9.4.0, 11.3.0), but should compile with most standard compilers.
  \item {\bf Run-time environment: } Tested on Ubuntu 20.04 and 22.04, but should broadly run on any Linux distribution.
  \item {\bf Hardware: } Running all simulations with 100,000 iterations for Row Hammer thresholds of 4800, 2400, and 1200 requires a single-core CPU.
  \item {\bf Metrics: } Attack Time (seconds and days). 
  \item {\bf Output: } Results shown in \cref{fig:attacktime_vs_round}.
  \item {\bf Experiments: } Instructions to run the experiments and parse the results are available in the README file.
  \item {\bf How much time is needed to complete experiments (approximately)?: } 3 minutes with a single-core Intel Xeon CPU.
  \item {\bf Publicly available?: } Yes.
  \item {\bf Archived (provide DOI)?: } \url{https://doi.org/10.5281/zenodo.7445036}
\end{itemize}
}
\noindent \subsubsection{Performance Evaluations}
{\small
\begin{itemize}
  \item {\bf Algorithm: } Implementation of Scale-SRS and RRS structures and operations in C. 
  \item {\bf Program: } The artifact assumes memory-access traces are available (filtered through an L1 and L2 cache model) for all of the benchmarks. This can be generated with any tracing tool (like Intel Pin~\cite{luk2005pin} v2.12). We tested the artifact with benchmarks from SPEC-2006, SPEC-2017, PARSEC, BIOBENCH, and GAP suites.
  \item {\bf Compilation: } Tested with gcc (version 11.3.0), but should compile with most standard compilers.
  \item {\bf Run-time environment: } Tested on Ubuntu 22.04, but should broadly run on any Linux distribution.
  \item {\bf Hardware: } Running all 78 benchmarks in parallel (78 simultaneous instances of the simulator) requires a CPU with a sufficient number of cores (64+) and memory (128GB+). 
  \item {\bf Metrics: } Normalized Performance (IPC). 
  \item {\bf Output: } Performance results shown in \cref{fig:perf4800}.
  \item {\bf Experiments: } Instructions to run the experiments and parse the results are available in the README file.
  \item {\bf How much time is needed to complete experiments (approximately)?: } 15 hours on Intel Xeon CPU if all 78 benchmarks are run in parallel (7-8 hours for baseline and RRS each on our system).
  \item {\bf Publicly available?: } Yes.
  \item {\bf Archived (provide DOI)?: } \url{https://doi.org/10.5281/zenodo.7445036}
\end{itemize}
}

%%%%%%%%%%%%%%%%%%%%%%%%%%%%%%%%%%%%%%%%%%%%%%%%%%%%%%%%%%%%%%%%%%%%%
\subsection{Access to the Artifact}
The code is available at \url{https://github.com/STAR-Laboratory/scale-srs} 
% \begin{itemize}
% \item  The code is available at \url{https://github.com/STAR-Laboratory/scale-srs} 
% \item The benchmark traces are available at \url{https://drive.google.com/file/d/1scEhit3nKWwnZwHiWLMBZ_lyNZzXoyzX/view?usp=sharing}
% \end{itemize}

\subsection{System Requirements and Dependencies}
\subsubsection{Requirements for Security Evaluations}

\begin{itemize}
    \item \textbf{Software Dependencies}: C++, Python3, g++ (tested to compile successfully with the version: 9.4.0 and 11.3.0), and Python3 Packages (pandas and matplotlib).
    \item \textbf{Hardware Dependencies}: A single-core CPU desktop/laptop will allow 100,000 iterations of Monte Carlo simulations in 1-3 minutes.
    \item \textbf{Data Dependencies}: Several input values, such as the number of attack rounds and the success probability of attack in a single refresh interval ($p_{k, T_S}$) in Equation~\ref{eqa:pkt}, are required to run the simulation. We generated these values following the equations in Section~\ref{security} and included the values in `scale-srs/security\_analysis/montecarlo-event/simscript/input'.
\end{itemize}

\subsubsection{Requirements for Performance Evaluations}

\begin{itemize}
    \item \textbf{Software Dependencies}: Perl (for scripts to run experiments and collate results) and gcc (tested to compile successfully with the version: 11.3.0).
    \item \textbf{Hardware Dependencies}: For running all the benchmarks, a CPU with lots of memory (128GB+) and cores (64+).
    \item \textbf{Trace Dependencies}: Our simulator requires traces of memory accesses for benchmarks (filtered through an L1 and L2 cache). We generate these traces using an Intel Pin~\cite{luk2005pin} (version 2.12). However, traces extracted in the format described at the end of the README file by any methodology (\textit{e.g.}, any Pin version) would be supported.
\end{itemize}

%%%%%%%%%%%%%%%%%%%%%%%%%%%%%%%%%%%%%%%%%%%%%%%%%%%%%%%%%%%%%%%%%%%%%
\subsection{Installation and Experiment Workflow}
\noindent \subsubsection{Security Evaluations}
The \texttt{run\_artifact.sh} in the scale-srs/security\_analysis/montecarlo-event folder performs all the steps required to compile, execute, collate results, and generate the results shown in \cref{fig:attacktime_vs_round}.
\begin{itemize}
\item \textbf{Compiles the code} using the Makefile in the scale-srs/security\_analysis/montecarlo-event folder.
\item \textbf{Executes the simulations} for all Row Hammer threshold values, first for \texttt{4800}, then for \texttt{2400}, and finally, for \texttt{1200}.
\item \textbf{Collates the results} for all benchmarks and provides the normalized performance.
\item \textbf{Reproduce the~\cref{fig:attacktime_vs_round}}.
\end{itemize}

\noindent \subsubsection{Performance Evaluations}
The \texttt{run\_artifact.sh} in the scale-srs/perf\_analysis folder performs all the steps required to compile, execute, collate results, and generate the results shown in \cref{fig:perf4800}.
\begin{itemize}
%\item \textbf{Downloads the trace files}.
\item \textbf{Compiles the code} using the Makefile in the scale-srs-/perf\_analysis/src folder.
\item \textbf{Executes the simulations} for all benchmarks in parallel (assuming the trace files are available), first for the \texttt{baseline}, then for the \texttt{Scale-SRS}, and finally, for the \texttt{RRS} configuration.
\item \textbf{Collates the results} for all benchmarks and provides the normalized performance.
\item \textbf{Reproduce the~\cref{fig:perf4800}}.
\end{itemize}

%%%%%%%%%%%%%%%%%%%%%%%%%%%%%%%%%%%%%%%%%%%%%%%%%%%%%%%%%%%%%%%%%%%%%
%\subsection{Experiment workflow}

%%%%%%%%%%%%%%%%%%%%%%%%%%%%%%%%%%%%%%%%%%%%%%%%%%%%%%%%%%%%%%%%%%%%%
\subsection{Evaluation and Expected Results}
\noindent \subsubsection{Security Evaluations}
The artifact provides the \texttt{get\_results\_4800.py, get\_results\_2400.py, and get\_results\_1200.py} files in the scale-srs/security\_analysis/montecarlo-event/simscript folder. This script allows the collation of the results, and the commands to collate the successful attack time of Juggernaut against RRS are provided in the \texttt{run\_artifact.sh} in the scale-srs/security\_analysis/montecarlo-event folder and the README file. After the completion of the \texttt{run\_artifact.sh}, the successful attack time for Row Hammer thresholds of 4800, 2400, and 1200 can be obtained as the \texttt{aggregate\_trh\_4800}, \texttt{aggregate\_trh\_2400}, and \texttt{aggregate\_trh\_1200} in the scale-srs/security\_analysis/montecarlo-event/results folder. Also, the regenerated \cref{fig:attacktime_vs_round} can be obtained as the \texttt{Figure6.pdf} file in the scale-srs/security\_analysis/montecarlo-event/graph folder. The sample results files for all of the used Row Hammer threshold values are provided in the scale-srs/security\_analysis/montecarlo-event/result folder.

\noindent \subsubsection{Performance Evaluations}
The artifact provides the \texttt{plot.sh} file in the scale-srs-/perf\_analysis/simscript folder. This script allows the collation of the results, and the commands to collate the IPC are provided in the \texttt{run\_artifact.sh} in the scale-srs/perf\_analysis folder and the README file. After the completion of the \texttt{run\_artifact.sh}, the normalized performance for all benchmarks can be obtained as the \texttt{data.csv} file in the scale-srs-/perf\_analysis/simscript folder. Also, the regenerated \cref{fig:perf4800} can be obtained as the \texttt{Figure14.pdf} file in the scale-srs-/perf\_analysis/graph folder. The sample results files for the baseline, Scale-SRS, and RRS configurations for all the benchmarks are provided in the scale-srs-/perf\_analysis/output folder of the artifact.

%%%%%%%%%%%%%%%%%%%%%%%%%%%%%%%%%%%%%%%%%%%%%%%%%%%%%%%%%%%%%%%%%%%%%%
%\subsection{Experiment customization}

%%%%%%%%%%%%%%%%%%%%%%%%%%%%%%%%%%%%%%%%%%%%%%%%%%%%%%%%%%%%%%%%%%%%%
%\subsection{Notes}

%%%%%%%%%%%%%%%%%%%%%%%%%%%%%%%%%%%%%%%%%%%%%%%%%%%%%%%%%%%%%%%%%%%%%
\subsection{Methodology}

Submission, reviewing and badging methodology:

\begin{itemize}
\small
  \item \url{https://ctuning.org/ae/reviewing.html}
  \item \url{http://cTuning.org/ae/submission-20201122.html}
  \item \url{http://cTuning.org/ae/reviewing-20201122.html}
\end{itemize}
}
% \newpage
\balance
\bibliographystyle{IEEEtranS}
\bibliography{citations}
\end{document}